\documentclass{ws-ijmpa}
\usepackage{physics}

\usepackage[caption=false,font=footnotesize]{subfig} 
\usepackage[super]{cite}
\usepackage{booktabs}
\usepackage{multirow}

\usepackage{float}
\usepackage{siunitx}
\usepackage{algorithm}
\usepackage{algpseudocode}

\usepackage{xcolor}
\usepackage[verbose,hypertexnames=false]{hyperref}
\hypersetup{colorlinks=false,allbordercolors=blue,pdfborderstyle={/S/U/W 1}}

\newcommand{\mysubcaption}[2]{\\\captionsetup[subfloat]{captionskip=-16pt}\subfloat[#1]{\makebox[\linewidth]{\label{#2}}}}

\begin{document}

\markboth{K.-I.~Ishikawa and S.~R.~Pandey}{Operator subspace based method for the extraction of higher energy levels}


%
\catchline{}{}{}{}{}
%

\title{Operator subspace based method for the extraction \\
       of higher energy levels in Lattice field theoretic systems}

\author{Ken-Ichi Ishikawa and Siddharth Ramesh Pandey}

\address{Graduate School of Advanced Science and Engineering, Hiroshima University,\\
         Higashi-Hiroshima, Hiroshima 739-8526, Japan\\
siddharth-pandey@hiroshima-u.ac.jp, ishikawa@theo.phys.sci.hiroshima-u.ac.jp}

\maketitle


\begin{abstract}
Given the importance of spectral analysis of lattice quantum field theory data, the advancement of techniques for the same remains
critically important. Therefore, we propose an algorithm that uses the first three
time slices of the correlation function matrix to construct the transfer matrix and extract its energy eigenvalues. Reduction of 
systematic error arising from truncating the operator basis is achieved by varying the operator subspace dimension and applying 
eigenvalue-variance extrapolation. Since the correlation functions in early time slices typically exhibit small statistical errors 
and retain a strong signal of higher-excited states, we expect our proposed method to perform well in extracting higher-excited-state 
energies. In this paper, we conduct two lattice Monte Carlo simulations for quantum mechanical systems, harmonic and anharmonic 
oscillators, to evaluate the efficiency of our proposed method for higher-excited-state energies. We compare the energy levels 
obtained using the standard GEVP method and our proposed method. We find that our method consistently outperforms 
the standard GEVP method in extracting intermediate excited states for both systems.
Because our method has access only to the first three time slices, it is not preferable to the GEVP method for extracting 
the ground-state energy in its current form.
\end{abstract}



\section{Introduction}

The spectral analysis extracting the energy levels of quantum theories is the most fundamental starting point to investigate properties of a target system.
The lattice quantum field theory (LQFT), which provides a non-perturbative method in analyzing quantum systems and is formulated on 
the Euclidean path-integral regularized on a lattice discretized Euclidean space-time, 
requires analysis of the Euclidean time dependence of correlation functions for the spectral analysis.

A correlation function matrix, which is evaluated via Monte Carlo simulations of the LQFT for a quantum system, behaves as 
\begin{align}
C_{ij}(t) &= \mel{0}{\hat{O}_i(t) \hat{O}_j^{\dagger}(0)}{0}
           = \sum_{n=0}^{\infty} A_{ij}^{(n)} e ^{-E_{n}t},
\label{eq:Cdefgen}
\end{align}
where $\hat{O}_i(t)$ are operators in the Euclidean Heisenberg picture and $t$ is the Euclidean time index (integers) on the lattice.
The energy levels $E_{n}$ and amplitudes $A_{ij}^{(n)}=\mel{0}{\hat{O}_i}{n}\mel{n}{\hat{O}_j}{0}$ with $\ket{n}$ the energy $n$-th eigenstate 
could be extracted from the time dependence of the correlation function matrix.
The energy and amplitude of the ground state $\ket{0}$ is usually accessible in a large time region
by observing an isolated exponential dumping as
\begin{align}
	C_{ij}(t) \to A_{ij}^{(0)} e ^{-E_{0}t}, \qquad (t \to \infty),
\label{eq:larget}
\end{align}
followed by a single exponential fitting. 
The extraction of excited states becomes difficult when we employ the same procedure with the multi-exponential fitting,
because the exponential factors, even with $E_n \ne E_m$ for any $n \ne m$, are not orthogonal 
yielding a large condition number in solving the fitting coefficients. 
The large condition number also makes the fitting more difficult, as it enhances the sensitivity of 
the fitting to the statistical error of Monte Carlo simulations in a large time region.
In this paper we propose a method to extract the excited state energy from the correlation function matrix of the LQFT.

One of the standard methods of extracting energy levels in the LQFT including lattice quantum chromodynamics (LQCD) 
is the generalized eigenvalue problem method (GEVP)\cite{michael-1985,Luscher:1990ck,Blossier:2008tx,Blossier:2009kd}.
This method basically shares the same philosophy of multi-exponential fitting, but it restructures 
the problem into solving the generalized eigenvalue problem to 
construct a better operator basis coupled to each state. 
The time dependence analysis is applied on each principal correlator which is obtained as the generalized eigenvalues.
After observing the isolation of single exponential behavior in each principal correlator the single exponential fittings is applied.
The fitting also suffers from the statistical error on the principal correlators, especially for excited states.

Recently, another method for the spectral analysis has been proposed and is becoming a standard method, 
which analyzes the eigenvalues of the transfer-matrix obtained from the correlation function matrix\cite{Wagman:2024rid,Hackett:2024xnx,Hackett:2024nbe,Chakraborty:2024exj,Ostmeyer:2024qgu,
Tsuji:2025zdn,Ostmeyer:2025igc}.
The time dependence of the correlation function matrix is translated into the power of 
the transfer-matrix on a state yielding a Krylov subspace to analyze the transfer-matrix.  
The method is also equivalent to an extension of the Prony method\cite{deProny1795,10.1063/1.1149581}, 
where the transfer-matrix is embedded in a form of Hankel matrix constructed from the correlation function matrix time series\cite{Ostmeyer:2024qgu,Chakraborty:2024exj}.
The implicit Lanczos algorithm, or explicit analysis on the transfer-matrix or Hankel matrix are
applied to obtain the spectrum of the transfer-matrix, which is the exponential of energies.

In this paper we focus on the extraction of higher excited states 
constrained by statistics and numerical precision involved in the Monte Carlo simulations.
The earlier works mainly focus on the ground and a few excited states.
The excited state signals involved in correlation functions evaluated through numerical
simulations quickly disappear because of the statistical error and of the lack of numerical precision.
When we apply existing methods for extracting energies, 
the systematic errors from the truncation of basis of Krylov or operators becomes more important.
The transfer-matrix based methods construct the Krylov subspace from the correlation function
by increasing the time range that suffer from large statistical errors in a later time region
yielding spurious eigenvalue problem in extracting excited state energies.

To address this, we explore the eigenvalue-variance extrapolation method\cite{Tsuji:2025zdn,imada-2000,kashima-2001} 
to extract higher-excited state energies using correlation function matrices at earlier time slices with a large operator basis.
We refer to this approach as the OGEVP method: the operator subspace based GEVP.
Using the data from earlier time slices, we can retain excited state signals with small statistical errors before 
they exponentially disappear.
In these constraints, we will investigate the method for higher excited states of 
quantum mechanical systems by changing the numerical precision, single and double precision.

This paper is organized as follows. 
In Section~2, we explain our proposed method, which applies eigenvalue analysis and eigenvalue-variance extrapolation
to the transfer-matrix derived from correlation function matrices with a large operator basis. We also present the systematic error of the method and the algorithm.
In Section~3, we present the simulations details along with presenting 
the results and subsequent discussion of the obtained results for two quantum mechanical systems; harmonic and anharmonic oscillators.
Section~4, we summarize our proposed method and finding in this paper along with the future prospects.

\section{Eigenvalue-variance extrapolation on Transfer-matrices}
As stated in the introduction, our proposed method combines the GEVP analysis and the eigenvalue-variance extrapolation method of transfer-matrices.
In this section, we first introduce the GEVP analysis on a transfer-matrix derived from a correlation function matrix by change of basis.
The application of the eigenvalue-variance extrapolation method is applied for minimizing the systematic error 
from the finite number of basis operators.
We then describe the whole algorithm of the OGEVP method.

The correlation function matrix to be investigated is shown in Eq.~\eqref{eq:Cdefgen}, which is obtained from a lattice simulation for a QM system.
We assume that the system has a transfer-matrix operator $\hat{T}$ and its Hamiltonian operator is 
implicitly defined though $\hat{T}=e^{-\hat{H}}$ on the lattice.
We also assume that the temporal lattice size is sufficiently large and can be treated as infinite. 
Thus, the correlation function matrix Eq.~\eqref{eq:Cdefgen} is evaluated with the ground state $\ket{0}$.
To be specific, we employ the number of operator basis to be $N_{\mathrm{OP}}$. 
The Euclidean time Heisenberg operator $\hat{O}_{i}(t)$ is defined by $\hat{O}_i(t) = \qty(\hat{T}^{\dag})^t \hat{O}_i\hat{T}^t$ 
with $\hat{O}_i$ in the Schr\"{o}dinger picture. 

For consecutive time slices $(t_0, t_0+1, t_0+2)$,
we can project the transfer-matrix operator $\hat{T}$ on a numerical matrix form 
using a set of states spanned by 
\begin{align}
S =\qty{\ket{\psi_i}\equiv T^{t_0}\hat{O}_i\ket{0}, i=1, 2, \dots, N_{\mathrm{OP}}}.
\label{eq:basisS}
\end{align}
This is not an orthonormal basis set, but the linear independence is usually assumed upon a choice of the operator set.
We also assume that the set exhausts the states of the target system in the limit of $N_{\mathrm{OP}}\to \infty$.

Using the basis set \eqref{eq:basisS} the correlation function matrix is written as
\begin{alignat}{3}
  C_{ij}(t_0)   &= \mel{0}{ \qty(\hat{T}^{\dag})^{t_0}\hat{O}_i \hat{T}^{t_0}\hat{O}^{\dag}_j}{0}     &{}={}& z^{t_0}  \braket{\psi_i}{\psi_j} &{}={}& z^{t_0} B_{ij},
\label{eq:Corr_at_t0}
\\
  C_{ij}(t_0+1) &= \mel{0}{ \qty(\hat{T}^{\dag})^{t_0+1}\hat{O}_i \hat{T}^{t_0+1}\hat{O}^{\dag}_j}{0} &{}={}& z^{t_0+1} \mel{\psi_i}{\hat{T}}{\psi_j} &{}={}& z^{t_0+1} T_{ij},
\label{eq:Corr_at_t1}
\\
  C_{ij}(t_0+2) &= \mel{0}{ \qty(\hat{T}^{\dag})^{t_0+2}\hat{O}_i \hat{T}^{t_0+2}\hat{O}^{\dag}_j}{0} &{}={}& z^{t_0+2} \mel{\psi_i}{\hat{T}^2}{\psi_j} &{}={}& z^{t_0+2} W_{ij},
\label{eq:Corr_at_t2}
\end{alignat}
where $z=e^{-E_0}$ is the ground state contribution associated with $\bra{0}\hat{T}^{\dag}=z\bra{0}$.
When $N_{\mathrm{OP}}$ is sufficiently large Eq.~\eqref{eq:Corr_at_t1} contains a good approximation of $\hat{T}$ projected on $S$.

Similarly to the transfer-matrix the GEVP method, the eigenvalue analysis on $T_{ij}$ proceeds
with solving the following generalized eigenvalue problem;
\begin{alignat}{2}
   C(t_0+1)  \vec{v}_{n} & = \lambda_n C(t_0) \vec{v}_n
      &{}\to{}&  T\vec{v}_{n} =  \bar{\lambda}_n B  \vec{v}_n,
\label{eq:Tinfdef}
\end{alignat}
with $\bar{\lambda}_n = e^{-(E_n-E_0)}$ at $N_{\mathrm{OP}}\to \infty$. 
The equation (\ref{eq:Corr_at_t2}) also has
\begin{alignat}{2}
   C(t_0+2)  \vec{v}_{n} & = \lambda_n C(t_0) \vec{v}_n
      &{}\to{}&  T^2\vec{v}_{n} =  \qty(\bar{\lambda}_n)^2 B  \vec{v}_n,
\label{eq:TTinfdef}
\end{alignat}
at $N_{\mathrm{OP}}\to \infty$. 
However, a finite $N_{\mathrm{OP}}$ causes systematic errors in the eigenvalue $\bar{\lambda}_n = e^{-(E_n-E_0)}$,
similar to the standard GEVP method, and it has been analyzed in Refs.~\citen{Blossier:2008tx,Blossier:2009kd}.
We also note that Eqs.~\eqref{eq:Tinfdef} and \eqref{eq:TTinfdef} do not share common eigenvectors $\vec{v}_n$ with
a finite operator basis.

The eigenvalue-variance extrapolation method focuses 
on the difference of Eqs.~\eqref{eq:Tinfdef} and \eqref{eq:TTinfdef} at a finite operator basis.
We expect that the difference between eigenvectors disappears by increasing $N_{\mathrm{OP}}$ and extrapolating it to infinity.
To achieve this, we first transform the generalized eigenvalue equation
\eqref{eq:Tinfdef} to the standard eigenvalue equation at a fixed finite $N_{\mathrm{OP}}$  with $ n<N_{\mathrm{OP}}$ ;
\begin{align}
  T\vec{v}_n &= \lambda_n \vec{v}_n,
\label{eq:change_of_basis}
\end{align}
where we substituted
  $T \Leftarrow  D^{-1/2}L^{\dagger}TLD^{-1/2}$ and
  $\vec{v}_n \Leftarrow  L^{\dagger}\vec{v}_n$ with 
  $L$ is the eigenvetor matrix of $B$ satisfying $BL=LD$ with diagonal matrix $D$.
We use the symbols $\lambda_n$ and $\vec{v}_n$ at a finite $N_{\mathrm{OP}}$, which are different from those at $N_{NOP}=\infty$.
The deviation of the eigensystem of Eq.~\eqref{eq:TTinfdef} from that of Eq.~\eqref{eq:Tinfdef} 
is monitored through the eigenvalue-variance $\Delta \lambda_n$ defined by
\begin{align}
\Delta \lambda_{n}& \equiv 
\dfrac{\vec{v}_n^{\dag} T_2 \vec{v}_n - \qty(\vec{v}_n^{\dag} T \vec{v}_n)^2}
      {\qty(\vec{v}_n^{\dag} T   \vec{v}_n)^2},
\label{eq:defdellam}
\\
T_2 &\equiv D^{-1/2}L^{\dagger}W LD^{-1/2}.
\label{eq:defT_2}
\end{align}
Because of the finite $N_{\mathrm{OP}}$, $T_2$ is not diagonalized by $\vec{v}_n$ indicating $\Delta \lambda_{n}\ne 0$.

The eigenvalue-variance extrapolation method extrapolates $(\lambda_n, \Delta \lambda_n)$ by increasing $N_{\mathrm{OP}}$.
In order to clarify the operator basis dimension, 
we use $(\lambda^{(N_{\mathrm{OP}})}_n, \Delta \lambda^{(N_{\mathrm{OP}})}_n)$ for the energy and variance obtained
with the operator basis dimension $N_{\mathrm{OP}}$. 
By changing $N_{\mathrm{OP}}$ from 1 to $N_{\mathrm{OPmax}}$, 
where $N_{\mathrm{OPmax}}$ a fixed maximum size of operator basis while keeping the computational cost in 
mind,
we can accumulate the data set 
\begin{align}
\qty{ \{ (\lambda^{(N_{\mathrm{OP}})}_n, \Delta \lambda^{(N_{\mathrm{OP}})}_n), n=0,1,\dots N_{\mathrm{OP}}-1\},
  N_{\mathrm{OP}}=1,\dots, N_{\mathrm{OPmax}} }.
\label{eq:all}
\end{align}
For a given energy level $n$ satisfying $0 \le n \le N_{\mathrm{OPmax}}-2$,
we can extrapolate $\lambda_n$ as a function of $\Delta \lambda_n$ in the limit $\Delta \lambda_n\to 0$
using the data set
\begin{align}
S_n \equiv \qty{ (\lambda^{(N_{\mathrm{OP}})}_n, \Delta \lambda^{(N_{\mathrm{OP}})}_n), N_{\mathrm{OP}}=n+1,\dots, N_{\mathrm{OPmax}} }.
\end{align}
Note that the data set $S_{N_{\mathrm{OPmax}}-1}$ cannot be used to perform the extrapolation $\Delta \lambda \to 0$
because it contains only one data.

With the detailed analysis on the systematic error of the exact eigenvalue $\bar{\lambda}_n$
the systematic error $\delta \lambda_n\equiv \bar{\lambda}_n-\lambda_n$ behaves as
\begin{align}
  \delta \lambda_n= -\dfrac{\bar{\lambda}_{n}}{2} \Delta\lambda_n + \epsilon \bar{\lambda}_{n},
  \label{eq:del_lam_var_rel}
\end{align}
where $\epsilon$ represents a small constant from the operator basis truncation effect~\cite{imada-2000,kashima-2001}.
Given Eq.~\eqref{eq:del_lam_var_rel}, $\lambda_n$ can be regarded as a function of $\Delta\lambda_n$:
\begin{align}
  \lambda_n(\Delta \lambda_n)  = \bar{\lambda}_n + \frac{-\bar{\lambda}_n}{2}\Delta \lambda_n + \epsilon \bar{\lambda}_{n}.
  \label{eq:del_lam_var_rel_diff_form}
\end{align}
The energy level $\tilde{E}_n$ at a given $(\lambda_n,\Delta \lambda_n)$ is defined by
\begin{align}
  \tilde{E}_n(\Delta \lambda_n)=-\ln{\left(\lambda_n(\Delta \lambda_n)\right)}.
  \label{eq:lamErel}
\end{align}

Using Eqs. \eqref{eq:lamErel} and \eqref{eq:del_lam_var_rel_diff_form} and taking $\Delta \lambda_n \to 0$ and expanding it around small $\epsilon$ the obtained energy is,
\begin{align}
\tilde{E}_n(0) = E_n-E_0 - \epsilon.
\label{eq:err_in_E2}
\end{align}
Therefore, the systematic error in extracting the energy level $E_n$ is minimized as
\begin{align}
\delta E_n = \order{\epsilon}.
\label{eq:err_in_E3}
\end{align}

\begin{algorithm}[t]
\caption{\small Extraction of energy levels using eigenvalue-variance extrapolation.}    
\label{tb:Algorithm}
\begin{algorithmic}[1]
\State{Create the correlation function matrix $C_{ij}(t) = \mel{0}{\hat{O}_i(t) \hat{O}_j^{\dagger}(0)}{0}$ 
       for  $i,j = 1,\dots N_{\mathrm{OP_{max}}}$ and $t=t_0, t_0+1, t_0+2$.}
\For{$N_{\mathrm{OP}}=1,\dots, N_{\mathrm{OP_{max}}}$ }
\State{Perform complete eigendecomposition of $C(t_0)$ as $C(t_0)L=DL$ where
       $L$ is the eigenvector matrix and $D$ is the diagonal eigenvalue matrix.  }
\State{Transform $C(t_0+1)$ and $C(t_0+2)$ for $T=D^{-1/2}L^{\dag}C(t_0+1)LD^{-1/2}$ 
       and $T_{2} = D^{-1/2}L^{\dag}C(t_0+2)LD^{-1/2}$}
\State{Evaluate the eigendecomposition for $T$ as $T V = V \Lambda$, 
       $V=(\vec{v}_0,\vec{v}_1,\dots,\vec{v}_{N_{\mathrm{OP}}-1})$, 
       $\Lambda = \mathrm{diag}\qty(\lambda_0,\lambda_1,\dots,\lambda_{N_{\mathrm{OP}}-1})$}
\State{Evaluate the eigenvalue variance $\Delta \lambda_n = \dfrac{\vec{v}_n^{\dag}T_2 \vec{v}_n - 
       \qty(\vec{v}_n^{\dag}T \vec{v}_n)^2}{\qty(\vec{v}_n^{\dag}T\vec{v}_n)^2}$ for $n=0,\dots,N_{\mathrm{OP}}-1$.}
\State{Store the data as $\qty{(\lambda_n^{(N_{\mathrm{OP}})}, \Delta \lambda_n^{(N_{\mathrm{OP}})}), n=0,\dots,N_{\mathrm{OP}}-1}$}
\EndFor
\State{Perform linear curve fitting on $\lambda_n$ as a linear function of $\Delta \lambda_n$ 
      using the data set 
      $\qty{ (\lambda^{(N_{\mathrm{OP}})}_n, \Delta \lambda^{(N_{\mathrm{OP}})}_n), N_{\mathrm{OP}}=n+1,\dots, N_{\mathrm{OPmax}} }$
      to get the energy level at zero variance $\lambda_n$ by taking the limit of $\Delta \lambda_n \to 0$
      for $n = 0, \dots,N_{\mathrm{OPmax}}-2$.}
\end{algorithmic}
\end{algorithm}

We present the algorithm to implement the proposed method in Alg.~\ref{tb:Algorithm}.
When implementing the algorithm we must choose the appropriate interpolating operators $\hat{O}_i$ 
  having a large overlap with the energies we wish to extract from the data.  
  After obtaining the $\lambda_n(0)$ for $\Delta \lambda=0$, corresponding energy, $\tilde{E}_n(0)$, 
 can be obtained by using Eq.~\eqref{eq:lamErel}.
 To estimate the statistical error of intermediate data and energies, 
 we repeat the algorithm exactly for the Jackknife~\cite{efron-1982,young-2015} data of correlation function matrices.

\section{Simulation Setup and Results}
In this section, we first describe the parameters involved in the simulations of harmonic and anharmonic oscillator systems. 
For the simulations we have used lattice Monte Carlo algorithms.
Next, the simulation results analyzed using both methods, our proposed and the standard GEVP method, are presented and discussed.

\begin{table}[t]
\centering
\tbl{Simulation parameters for the harmonic and anharmonic oscillator systems.\label{tab:simulation_parameters}}
{\begin{tabular}{@{}lcc@{}}
  \toprule
    \textbf{Quantity} & \textbf{Harmonic Oscillator} & \textbf{Anharmonic Oscillator} \\
  \colrule
    Parameters of the action  & $m=1,\;\omega = 1$            & $m=1,\;f=1,\;\lambda=1$ \\
     $(a,aN_T)$               & $(0.01,50),\;(0.02,50),$ & $(0.01,160),\;(0.02,80),$    \\
    			  & $\;(0.04,50)$            & $\;(0.04,40)$                 \\
    Simulation algorithm      & Heat-Bath            & Hybrid Monte-Carlo \\
    Thermalization sweeps     & $100000$             & $1000$ \\
    Total number of sweeps    & $60000000$           & $1500000$ \\
    Sweeps stored after every & $1000$               & $100$ \\
    Raw data samples          & $60000$              & $15000$ \\
    Bin size                  & $100$                & $100$ \\
    Binned samples            & $600$                & $150$ \\
  \botrule
\end{tabular}}
\end{table}

\subsection{Simulation setup}
The systems we studied are quantum mechanical harmonic and anharmonic oscillator systems. 
Since the actions for both systems in Eqs.~\eqref{eq:Harmaction} and 
\eqref{eq:anHarmaction} are both invariant under parity transformations $x \to -x$, 
the odd and even parity channels are disjoint. We focus on the odd-parity channel for both systems and 
use the correlation function matrix defined as,
\begin{align}
 C_{mn}(\tau)= \dfrac{1}{N_{T}} \sum_{t=0}^{N_T-1}\expval{ \hat{x}^{2m+1}(t+\tau) \hat{x}^{2n+1}(t) },
\label{eq:Corrdefodd}
\end{align}
where, $N_T$ is the time extent of the system and $m,n \in \mathbb{N}\cup\{0\}$.
We employ the operator basis: $\qty{\hat{O}_i(t),i=1,\dots,6}=\qty{ \hat{x}(t),\hat{x}^3(t)\dots, \hat{x}^{11}}$ 
with $N_{\mathrm{OPmax}}=6$.

The Euclidean action~\cite{creutz-1981} used to generate samples for harmonic oscillator is,
\begin{align}
	S(x)=a \sum_{i=0}^{N_{T}-1}\left\{ \frac{m}{2}\left(\frac{x_{i+1}-x_{i}}{a}\right)^2 + \frac{m\omega^2 x_{i}^2}{2}\right\},
\label{eq:Harmaction}
\end{align}
and for anharmonic oscillator~\cite{creutz-1981} is,
\begin{align}
S(x)=a \sum_{i=0}^{N_{T}-1}\left\{ \frac{m}{2}\left(\frac{x_{i+1}-x_{i}}{a}\right)^2 + \frac{\lambda}{2} \left(x_{i}^2-f^2\right) \right\}.
\label{eq:anHarmaction}
\end{align}
The data was generated in both single precision (denoted by subscripts $sp$) and double precision (denoted by subscripts $dp$). Averaging of single precision data
was done in single precision and averaging of double precision data was done in double precision.
The statistical analysis of the both data both done in 
double precision. 
We list the parameters of the simulations in Table~\ref{tab:simulation_parameters}.
The lattice spacing $a$ and the physical volume $aN_T$ are chosen to obtain 
the continuum limit in sufficiently large temporal extent $e^{-E_0 T} \ll 1$.
The statistical errors are evaluated with the Jackknife method~\cite{efron-1982,young-2015} 
after binning data. 
The binning size is obtained by observing the auto-correlation time of correlation function matrices.

\subsection{Results}

In this subsection we will discuss in detail the results obtained by applying the GEVP method 
and our proposed method on the correlation function matrix Eq.~\eqref{eq:Corrdefodd}.
We will be targeting up to the energy level $E_9$ for both harmonic and anharmonic systems. 
As the variance extrapolation method needs at least two data points 
$(\lambda,\Delta \lambda)$ with different operator dimension,
we can apply the OGEVP method up to $E_9$ ($\because  N_{\mathrm{OPmax}}=6$), where
two data points of $(\lambda_9,\Delta \lambda_9)$ are obtained with two operator basis sets:
 $\qty{\hat{x}(t),\hat{x}^3(t),\dots, \hat{x}^9(t)}$ (5 dimensions) and
 $\qty{\hat{x}(t),\hat{x}^3(t),\dots, \hat{x}^{11}(t)}$ (6 dimensions).
For comparison, we fix the dimension of correlation function matrix as $5$ in the case of the GEVP.
We set the starting time at $t_0=0$ for both the GEVP method and the OGEVP method.

\begin{table}[t]
 \tbl{Harmonic oscillator energies analyzed with the GEVP and OGEVP methods.\label{tb:har results}}
{\begin{tabular}{@{}lcccccc@{}}
  \toprule
     & $E^{GEVP}_{sp}$ & $E^{GEVP}_{dp}$ & $E^{OGEVP}_{sp}$ & $E^{OGEVP}_{dp}$ &  $E_{r}$  \\
  \colrule
       $E_9$ & $10(1)$      & $11(3)$       & $9(7)$        & $10.56(96)$    & $9$ \\
       $E_7$ & $7.39(36)$   & $7.92(47)$    & $7.20(20)$    & $7.34(35)$     & $7$ \\
       $E_5$ & $5.025(51)$  & $5.126(77)$   & $4.971(73)$   & $5.050(82)$    & $5$ \\
       $E_3$ & $2.989(13)$  & $3.002(13)$   & $2.988(15)$   & $3.015(14)$    & $3$ \\
       $E_1$ & $0.9989(22)$ & $0.9999(20)$  & $1.0016(27)$  & $1.0008(23)$   & $1$ \\
  \botrule
  \end{tabular}}
\end{table}

\begin{table}[t]
\tbl{Same as Table~\ref{tb:har results}, but for the anharmonic oscillator.\label{tb2}}
{\begin{tabular}{@{}lcccccc@{}}
  \toprule
   & $E^{GEVP}_{sp}$ & $E^{GEVP}_{dp}$ & $E^{OGEVP}_{sp}$ & $E^{OGEVP}_{dp}$ &  $E_{r}$  \\
  \colrule
    $E_9$ & ---          & $24(3)$        & ---           & $20(11)$       & $18.79604508$ \\
    $E_7$ & $14.14(66)$  & $13(2)$        & $13.54(29)$   & $13.16(59)$    & $13.296081$   \\
    $E_5$ & $8.439(86)$  & $8.35(23)$     & $8.282(62)$   & $8.331(69)$    & $8.348699$    \\
    $E_3$ & $4.107(24)$  & $4.1108(76)$   & $4.107(18)$   & $4.103(30)$    & $4.097541$    \\
    $E_1$ & $0.7879(13)$ & $0.78802(90)$  & $0.7928(30)$  & $0.7940(48)$   & $0.787621$    \\
  \botrule
\end{tabular}}
\end{table}

We first present the energies obtained for both harmonic and anharmonic oscillators after 
taking the continuum limit in Tabs.~\ref{tb:har results} and \ref{tb2}.
The energies are compared with $E_r$: solutions of the Schr\"{o}dinger equation obtained analytically or numerically.
~\ref{appxA} stores the $\lambda$ and $\Delta \lambda$ data and corresponding plots for anharmonic oscillator. It also stores lattice size wise energy data
for both harmonic and anharmonic oscillator.
In the following, we explain the details of these results for the harmonic oscillator 
and then the anharmonic oscillator.

\begin{figure}[h]
\centering
  \begin{minipage}{0.49\textwidth}
    \centering
    \includegraphics[width=\linewidth]{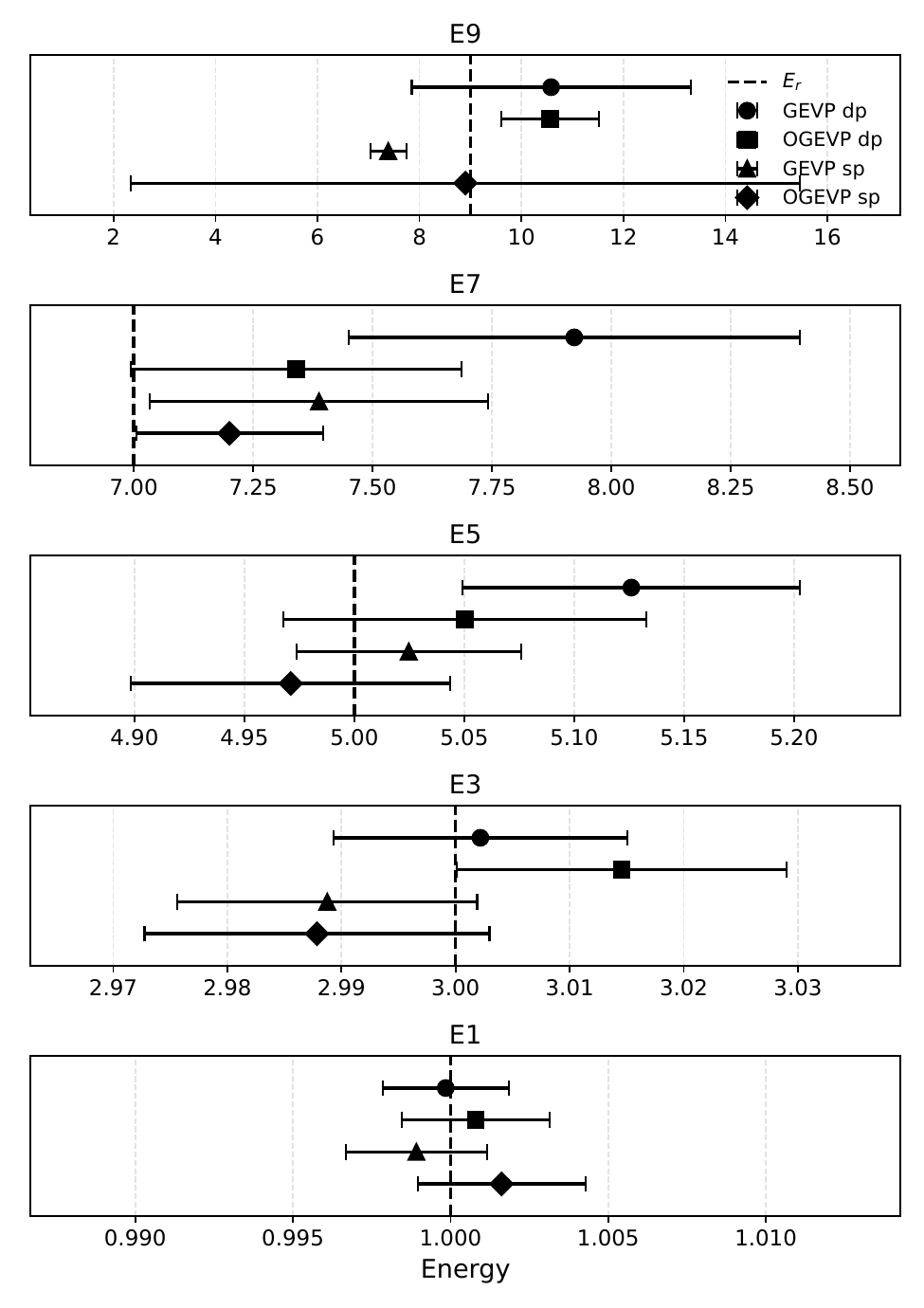}
    \mysubcaption{Energy comparison}{fig:harmcomp}
  \end{minipage}
\hfill
  \begin{minipage}{0.49\textwidth}
    \centering
    \includegraphics[width=\linewidth]{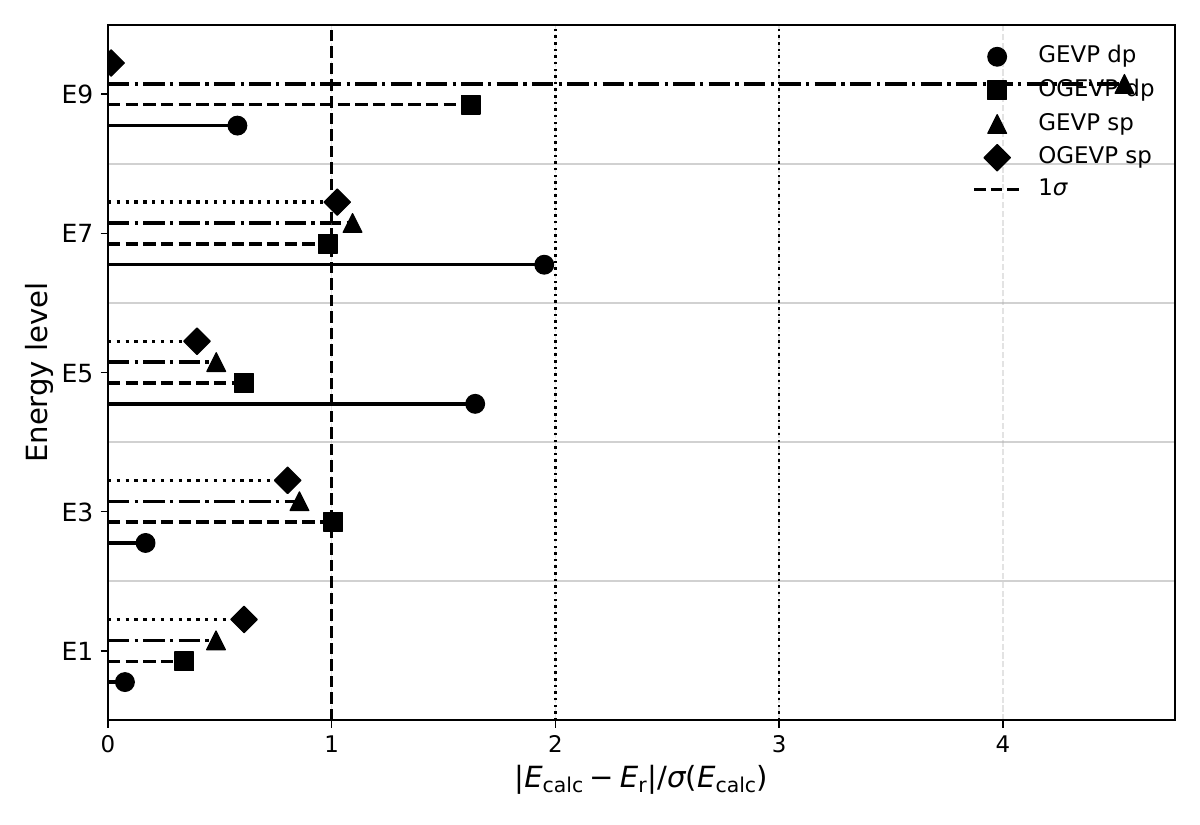}
    \mysubcaption{Pull plot}{fig:harmpull}
  \end{minipage}
  \caption{(a) Comparison of the harmonic oscillator energies obtained from different methods and different precision data sets.
           (b) Pull plot showing the deviation of the extracted energies from the analytic values $E_r$ in units of the statistical uncertainty. 
           The corresponding numerical values are given in Table~\ref{tb:har results}.}
  \label{fig:EnergyComparisonPullHarm}
\end{figure}

\paragraph{Harmonic oscillator:}
All results are compared with the analytic one as shown in Fig.~\ref{fig:EnergyComparisonPullHarm}.
The left-hand panel shows the energies obtained after the continuum limit in both types of numerical precision and analysis methods.
The right-hand panel displays the deviation from the analytic energy $E_r$ for visibility.
Given that the harmonic oscillator is an analytically solvable system
and the states coupled to the given operator set is explicitly fixed, 
the correlation function matrix with $N_\mathrm{OP} = 5$ only contains signals from $E_1,\dots, E_9$ states.
The higher excited states do not contaminate the signal. 
Thus, we observe that there is no significant difference in the energies obtained from 
the standard GEVP method and our proposed method as displayed in Tab.~\ref{tb:har results} and Fig.~\ref{fig:EnergyComparisonPullHarm}.
We can see a tendency that the statistical error becomes large for excited energies.
This behavior originates from the statistical nature of the correlation function matrix.
Both methods solve eigenvalue equations for the correlation function matrix and
the higher excited state energies correspond to the small eigenvalues.
The statistical error dominates more for smaller eigenvalues 
yielding larger statistical errors for corresponding exited energies.
The energies up to $E_7$ are consistent with the analytic ones within 2$\sigma$.
The highest energies $E_9$ are affected by large statistical error 
signaling the failure of stably solving generalized eigenvalue equations 
for the larger subspace dimension size in both methods.

\begin{figure}[t]
  \centering
  \begin{minipage}{0.49\textwidth}
    \centering
    \includegraphics[width=\linewidth]{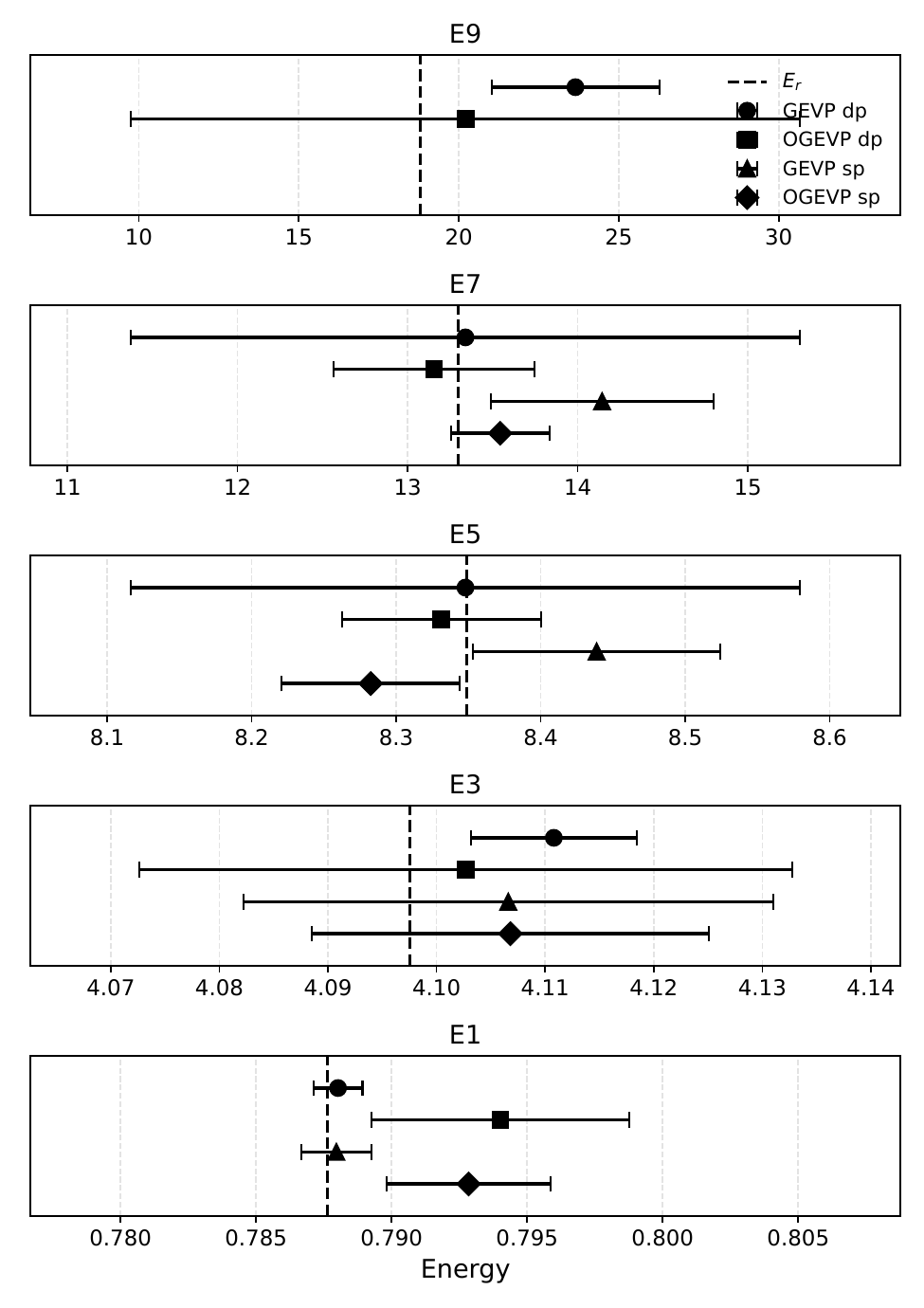}
    \mysubcaption{Energy comparison}{fig:anharmcomp}
  \end{minipage}
\hfill
  \begin{minipage}{0.49\textwidth}
    \centering
    \includegraphics[width=\linewidth]{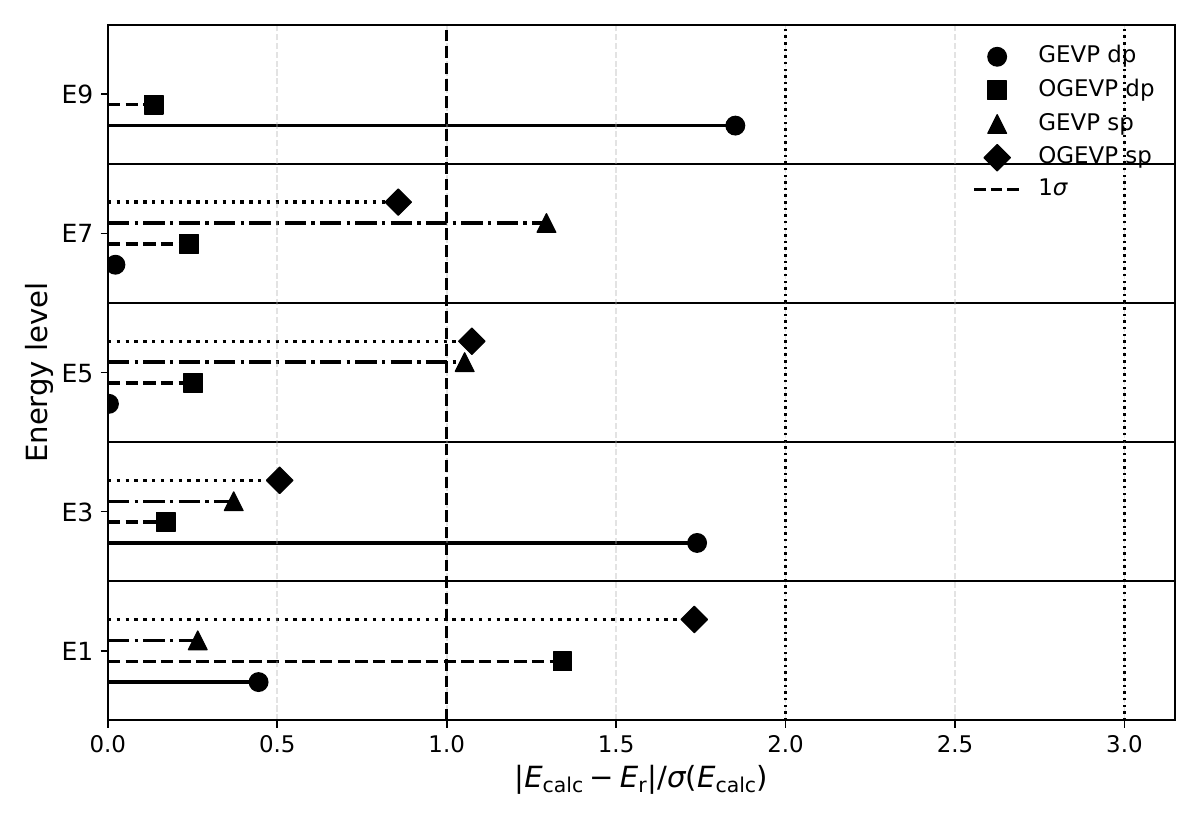}
    \mysubcaption{Pull plot}{fig:anharmpull}
  \end{minipage}
   \caption{Same as Figure~\ref{fig:EnergyComparisonPullHarm}, but for the anharmonic oscillator.
            The corresponding numerical values are listed in Table~\ref{tb2}.}
  \label{fig:EnergyComparisonPullAnharm}
\end{figure}

\ \\
\paragraph{Anharmonic oscillator:}
In the anharmonic oscillator system all the energies contaminate all the other channels with the operator basis we employ.
This makes the spectrum non-trivial and is an ideal place to compare the OGEVP method with 
the standard GEVP method.
Table~\ref{tb2} shows the energies obtained for anharmonic oscillator.
Figure~\ref{fig:EnergyComparisonPullAnharm} displays the energies compared with $E_r$.

Focusing on the ground state in the odd-channel $E_1$, 
we observe that
the GEVP method resolves the ground-state energy up to two- and three-decimal places for single and double precision data respectively with a small statistical error.
On the other hand, our proposed method fails to resolve the two-decimal places of the ground-state energy compared to $E_{r}$, though they stay 
within a rather large statistical error: $< 2\sigma$. This can be also observed in Fig.~\ref{fig:EnergyComparisonPullAnharm}.
This discrepancy between the GEVP and the OGEVP methods comes from the fact that 
the OGEVP method uses only three time slices while the GEVP method can access large time region.
The signal of the principal correlator of the GEVP method is dominated by the ground state with better statistics.
Figure~\ref{fig:EFF10p01},for example, shows the effective mass plot of the principal correlator with the GEVP at $a=0.01$,
where the gray-band represents the fitted energy in the fit range with error band. 
On the other hand, the OGEVP method evaluates the eigenvalue-variance pairs $(\lambda,\Delta \lambda)$ for 
a set of operator basis from the correlation function matrix at first three slices. 
Figure~\ref{fig:LAM1OGEVP0p01} shows the variance-extrapolation in the case of $a=0.01$.
The statistical error of $\lambda_0$ for each data point is consistent with those of the eigenvalue obtained 
with the GEVP at the corresponding time slices because they share the same correlation function matrix.
 However, the variance extrapolation, $\Delta \lambda \to 0$, 
 causes a rather larger error as seen in Fig.~\ref{fig:LAM1OGEVP0p01},
 where the horizontal error of $\Delta \lambda$ affects the vertical error significantly.
 We also note that there are several data points with $\Delta \lambda < 0$, 
 which is allowed as it is not statistical variance, since we have used the same basis 
 $v_n$ for both $T$ and $T_{2}$ matrices.

\begin{figure}[t]
  \centering
  \begin{minipage}[b]{0.49\textwidth}
    \centering
    \includegraphics[width=\linewidth]{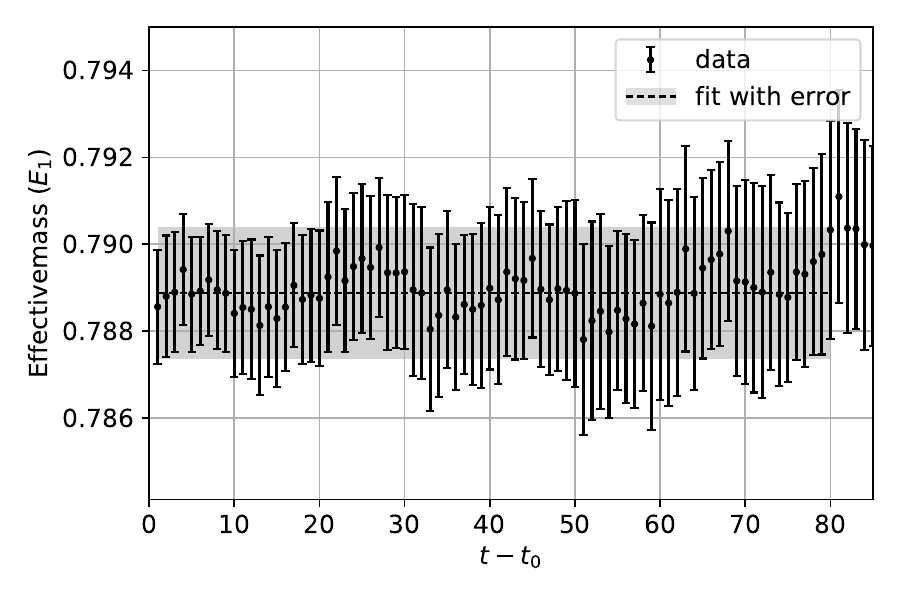}
    \mysubcaption{Effective mass plot for $E_1$.}{fig:EFF10p01}
  \end{minipage}
\hfill
  \begin{minipage}[b]{0.49\textwidth}
    \centering
    \includegraphics[width=\linewidth]{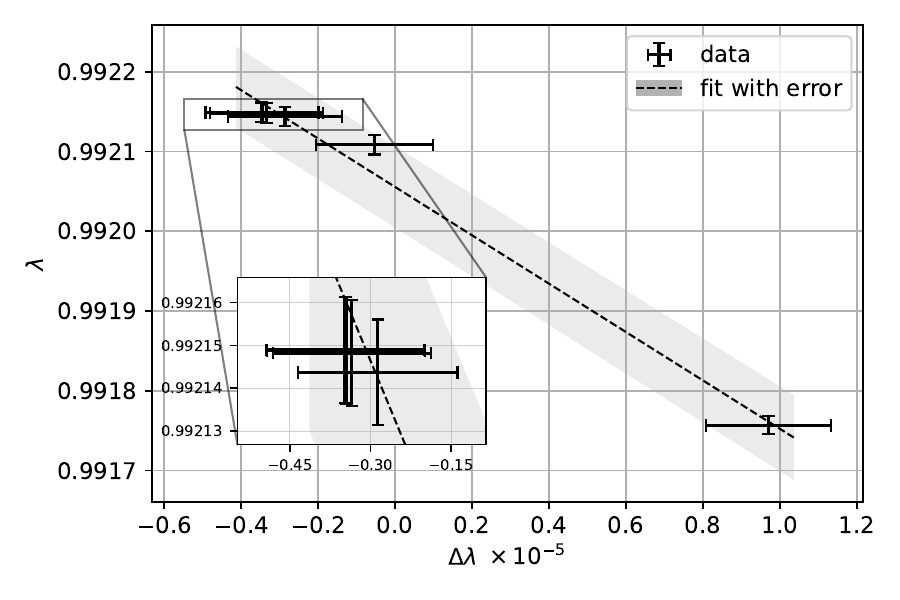}
    \mysubcaption{$\lambda$ vs $\Delta \lambda$ plot for $E_1$ excited state.}{fig:LAM1OGEVP0p01}
  \end{minipage}
\hfill
  \begin{minipage}[b]{0.49\textwidth}
    \centering
    \includegraphics[width=\linewidth]{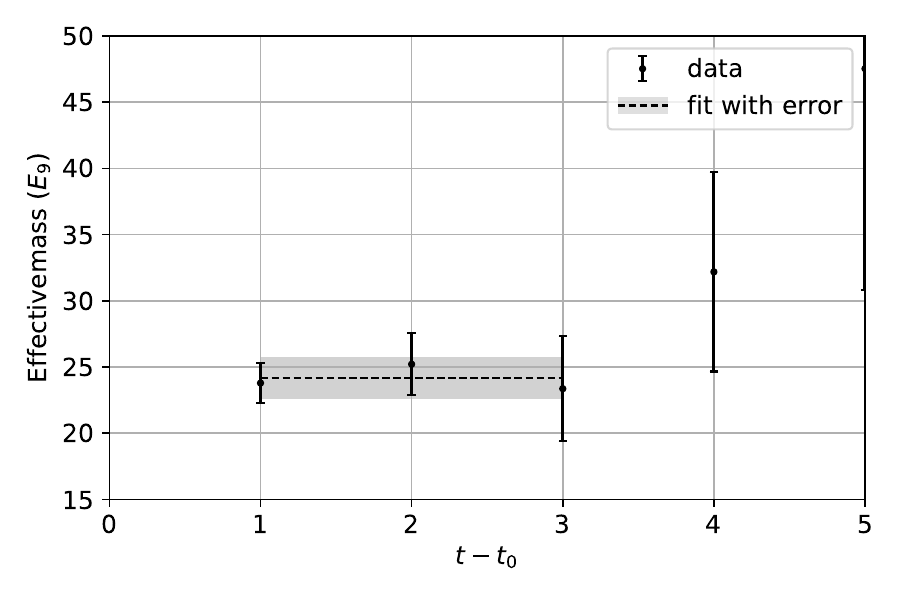}
    \mysubcaption{Effective mass plot for $E_9$.}{fig:EFF90p01}
  \end{minipage}
\hfill
  \begin{minipage}[b]{0.49\textwidth}
    \centering
    \includegraphics[width=\linewidth]{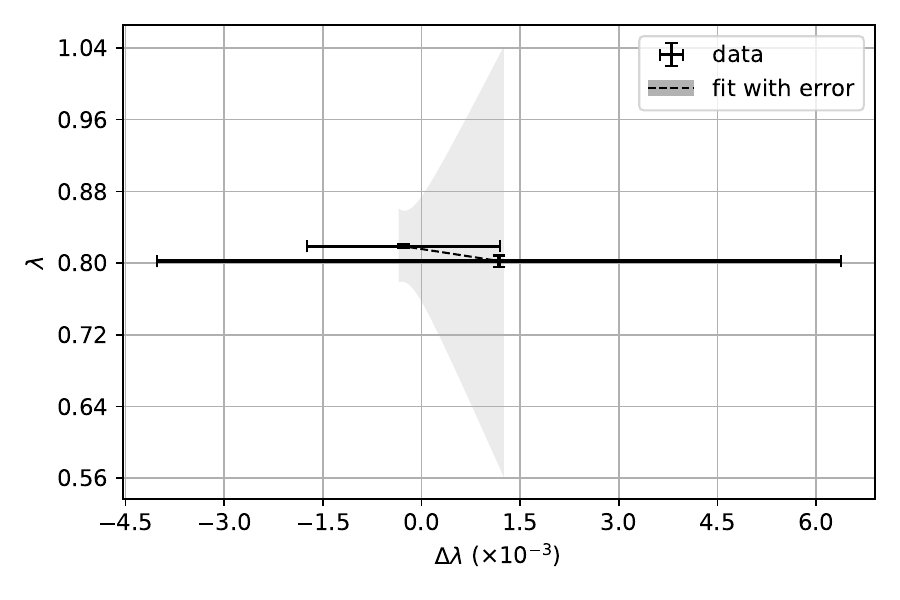}
    \mysubcaption{$\lambda$ vs $\Delta \lambda$ plot for $E_9$ excited state.}{fig:LAM9OGEVP0p01}
  \end{minipage}
  \caption{Comparison of the Effective mass plots and $\lambda$ vs $\Delta\lambda$ plots for the $E_1$ and $E_9$ excited states at lattice spacings $a=0.01$.
           In the effective mass plots the central dotted line is the energy obtained by fitting the principle correlator on the region denoted by the horizontal
           extent of the band.}
  \label{fig:E1andE9}
\end{figure}

For the $E_3$ level the results from the GEVP and the OGEVP method are numerically equivalent. 
We notice that for all the excited energy levels the double-precision data result from the OGEVP method is 
consistent to $0.5\sigma$ as shown in Fig.~\ref{fig:EnergyComparisonPullAnharm}.
In the case of $E_5$ level we observe that only our proposed method in double-precision is able to resolve
the energy level to 2 decimal places. 
Moreover, for $E_7$ our proposed method is able to correctly resolve up to 2 decimal places for both types of precision, 
while the GEVP method fails to resolve the energy as given in Tab.~\ref{tb2}, showcasing reliable signal extraction even in single-precision data for the OGEVP method.

Looking at $E_9$ the difference becomes unclear because of the large errors
for both methods, as seen in Figs.~\ref{fig:EFF90p01} and ~\ref{fig:LAM9OGEVP0p01}.
The signal becomes worse in both methods.
The variance extrapolation $\Delta \lambda \to 0$ using
two data points at $N_{\mathrm{OP}}=N_{\mathrm{OPmax}}-1$ and $N_{\mathrm{OPmax}}$
is not enough to resolve the signal in the presence of large statistical noise.
We can see that the single-precision case fails to extract the highest energy $E_9$ in both methods.
A naive reason for this is simply loss of signal due to the small dynamical range of the single-precision number format,
the excited state energy quickly falls below the numerical range responsible to the single-precision.
Since $E_{9}-E_{0}\sim 19$, the corresponding eigenvalue of the $T$ matrix in Eq.~\eqref{eq:change_of_basis} 
is $\exp\qty(-\qty(E_{9}-E_0)) \approx \order{10^{-9}}$,
which is actually below the limit of single precision numbers.
Thus, in conclusion the OGEVP method outperforms the standard GEVP 
when we can access enough data points for the variance extrapolation for excited states.

\begin{figure}[t]
\centering
  \begin{minipage}[b]{0.49\textwidth}
    \centering
    \includegraphics[width=\linewidth]{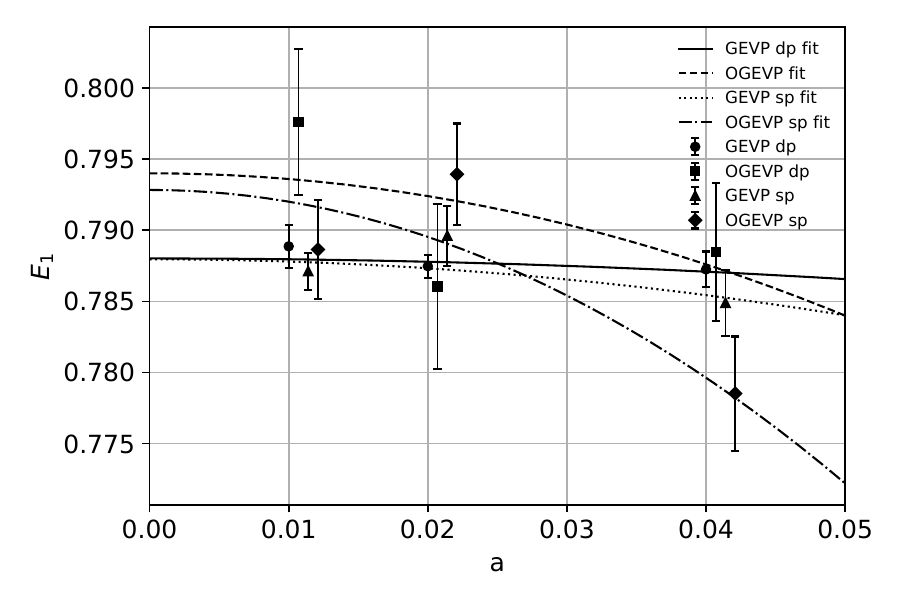}
    \mysubcaption{$E_1$}{fig:Eanharm_1}
  \end{minipage}
  \hfill
  \begin{minipage}[b]{0.49\textwidth}
    \centering
    \includegraphics[width=\linewidth]{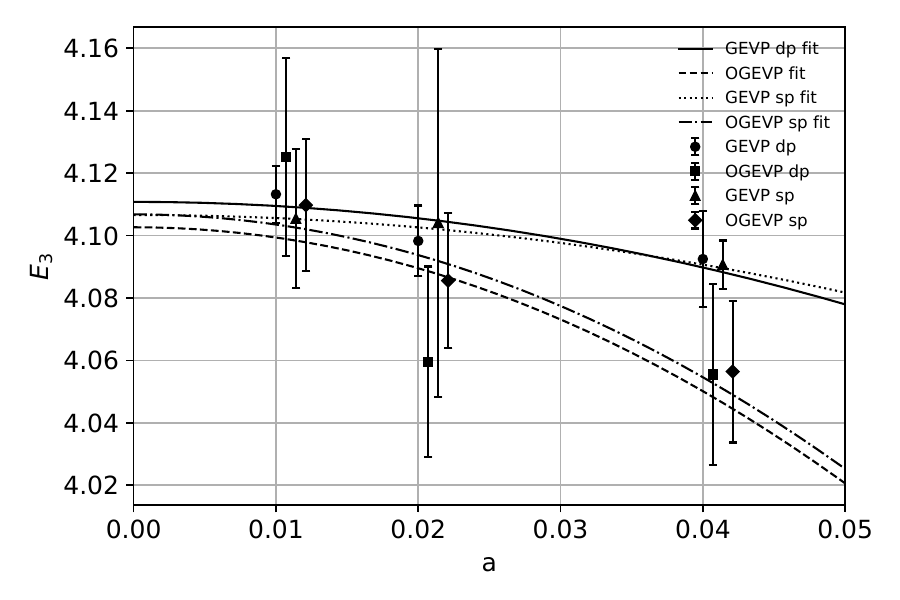}
    \mysubcaption{$E_3$}{fig:Eanharm_3}
  \end{minipage}
  \hfill
  \begin{minipage}[b]{0.49\textwidth}
    \centering
    \includegraphics[width=\linewidth]{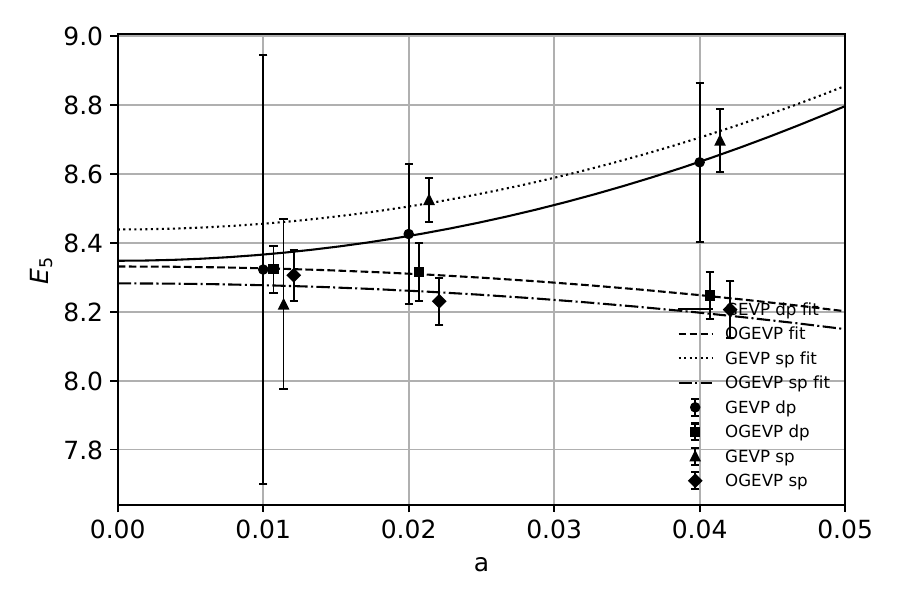}
    \mysubcaption{$E_5$}{fig:Eanharm_5}
  \end{minipage}
  \hfill
  \begin{minipage}[b]{0.49\textwidth}
    \centering
    \includegraphics[width=\linewidth]{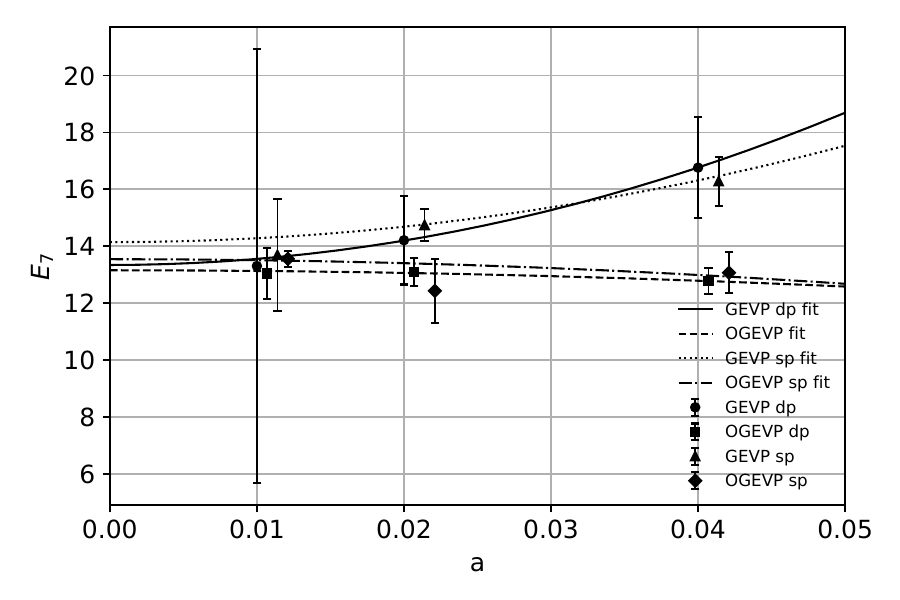}
    \mysubcaption{$E_7$}{fig:Eanharm_7}
  \end{minipage}
  \hfill
  \begin{minipage}[b]{0.49\textwidth}
    \centering
    \includegraphics[width=\linewidth]{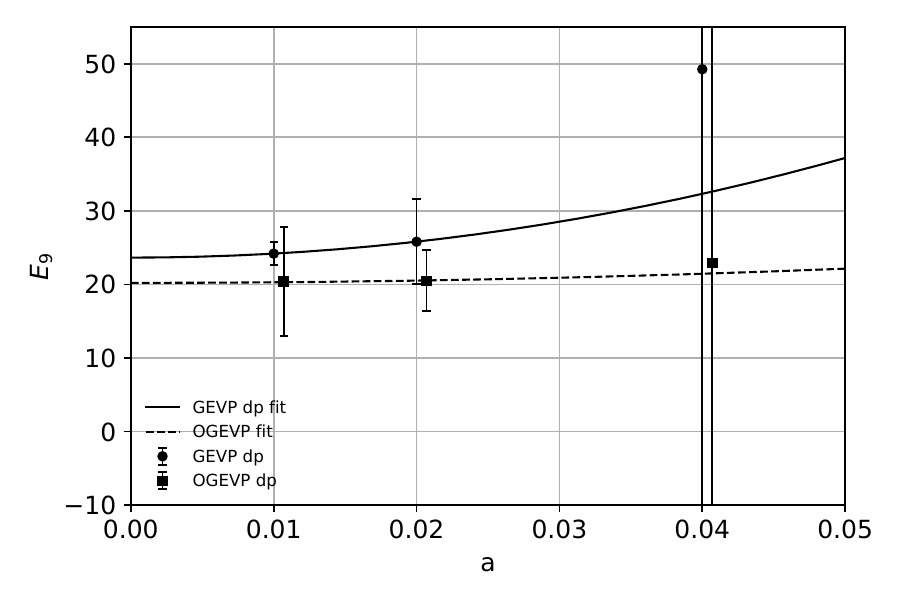}
    \mysubcaption{$E_9$}{fig:Eanharm_9}
  \end{minipage}
  \caption{Continuum limit plots for $E_1, \dots, E_9$ for anharmonic oscillator.}
  \label{fig:fivepanelENganharm}
\end{figure}

We show the lattice spacing dependence of the energies in Fig.~\ref{fig:fivepanelENganharm}.
For the lower state energies $E_1$ and $E_3$, 
the both results are consistent within $2\sigma$ at each finite $a$ 
(Figs.~\ref{fig:Eanharm_1} and \ref{fig:Eanharm_3}),
while the results for the excited state energies $E_5$ and $E_7$ 
deviate at $a=0.04$ (Figs.~\ref{fig:Eanharm_5} and \ref{fig:Eanharm_7}).
We speculate that this behavior is due to the different nature of systematic error between the GEVP and the OGEVP method. 
The systematic error for the OGEVP method is $\order{\epsilon}$ as 
in Eqs.~\eqref{eq:del_lam_var_rel} and \eqref{eq:err_in_E2}.
In the limit of $N_{\mathrm{OPmax}}\to\infty$ 
the $\order{\epsilon}$ term of Eq.~\eqref{eq:del_lam_var_rel} 
vanishes resulting $\Delta \lambda \to 0$. 
However, we observed $\Delta \lambda <0$
as an effect of insufficient statistics and 
due to our finite maximum operator dimension $N_{\mathrm{OPmax}}$ leading $\order{\epsilon}$ systematic error.
The eigenvalue-variance extrapolation method already takes this effect into the error of the results.
Meanwhile, for the GEVP method all the data points in the Fig.~\ref{fig:fivepanelENganharm}
correspond to the energy obtained by fitting the principal correlator 
over fixed plateau ranges from effective mass plots.
As we have not surveyed possible systematic errors,
such as $\order{e^{-(E_{N_{\mathrm{OP}}}-E_n)(t-t_0)}}$~\cite{Blossier:2009kd},
by varying $t_0$, fitting range, and the operator basis dimension in the GEVP method
(for practical surveys on the systematics of the GEVP method see Refs.~\citen{Blossier:2009kd,bulava-2012,bernardoni-2012}).
Therefore, the errors are underestimated for the GEVP method.
We also note that the $a$ dependence in the case of harmonic oscillator,
the results are perfectly consistent each other, thanks to the best choice of the operator basis
(the data are shown in Tab.~\ref{tab:precision_comparison_harm} of~\ref{appxA}).

\section{Summary and Future Prospects}
In this paper, we have presented our algorithm for the spectral analysis of 
the lattice field theoretic data. 
Our proposed method extends the application of the eigenvalue-variance extrapolation 
for the extraction of higher-excited states beyond it's established application to the ground state\cite{}.
Our proposed method focus on 
the correlation functions at the first three time slices 
having large overlaps to exited states with small statistical error.
And combined with the eigenvalue-variance extrapolation\cite{Tsuji:2025zdn,imada-2000,kashima-2001} with 
varying the dimension of operator basis for extracting excited state energies.

We have tested our proposed method to two quantum mechanical systems, harmonic and anharmonic oscillators.
Our method has been compared with the standard GEVP method.
Given the evidence in the previous sections,
we conclude that 
the OGEVP method successfully extract the intermediate excited state energies;
thanks to the small statistical error in first three time slices 
and minimization of the systematic error from the finite dimension of the operator basis.
Moreover, the OGEVP method outperforms the standard GEVP method for the excited states with a better resolution.
Having access to only three time slices makes our proposed method not suitable over the GEVP method for ground state energy extraction.
The combination of the eigenvalue-variance extrapolation and TGEVP is seemed 
to be the best for the ground state energy extraction~\cite{Tsuji:2025zdn}.

In the future, we would like to extend our proposed method beyond just 
the first three time slices of the correlation function matrix.
There are two motivations for the extension; 
one is to extract the ground-state energy more reliably,
and the second is to generalize our proposed method for improved lattice actions.
By utilizing the time translation invariance of the correlation function matrix
and varying $t_0 \in \mathbb{N}\cup\{0\}$,
we could extract the ground-state energy more reliably
using the all data $(\lambda_n(t_0), \Delta \lambda(t_0))$
for the variance extrapolation $\Delta \lambda \to 0$ to the former reason.
In this paper we have employed simple lattice quantum mechanical actions having nearest neighbor coupling.
For generic lattice actions with non-nearest neighbor interactions, 
the transfer-matrix extracted from the correlation functions at successive time 
slices may not simply be related to the transfer-matrix in the continuum-limit.
In this case the first three time slices are not suitable with our proposed method. 
A generalization of the time separation is needed to the latter reason.

\section*{Acknowledgments}
S.R.P acknowledges financial support from the Japanese Government (MEXT) Scholarship. We would like to thank Sinya Aoki and Alessandro Mariani for their insightful comments at 
Frontiers of Lattice Fermions Workshop 2026.


\section*{ORCID}
\noindent Siddharth Ramesh Pandey - \url{https://orcid.org/0009-0007-1612-9326}

\noindent Ken-Ichi Ishikawa - \url{https://orcid.org/0000-0001-6425-1894}

\appendix

\section{Data from the simulations}
\label{appxA}
In this appendix we collect all the energy eigenvalues from the GEVP and our proposed methods
at each lattice spacing. 
For the anharmonic oscillator, data for $\lambda$ and $\Delta\lambda$ and 
figures for the eigenvalue-variance extrapolation at each lattice spacing are shown.

Tables~\ref{tab:precision_comparison_harm} and \ref{tab:precision_comparison_aharm} show
the energies obtained from the GEVP method and the OGEVP method at lattice spacing $a \in \{0.01,0.02,0.04\}$
for both types of precision for harmonic and anharmonic oscillators.

\begin{figure}[p]
\centering
  \begin{minipage}{0.49\textwidth}
    \centering
    \includegraphics[width=\linewidth]{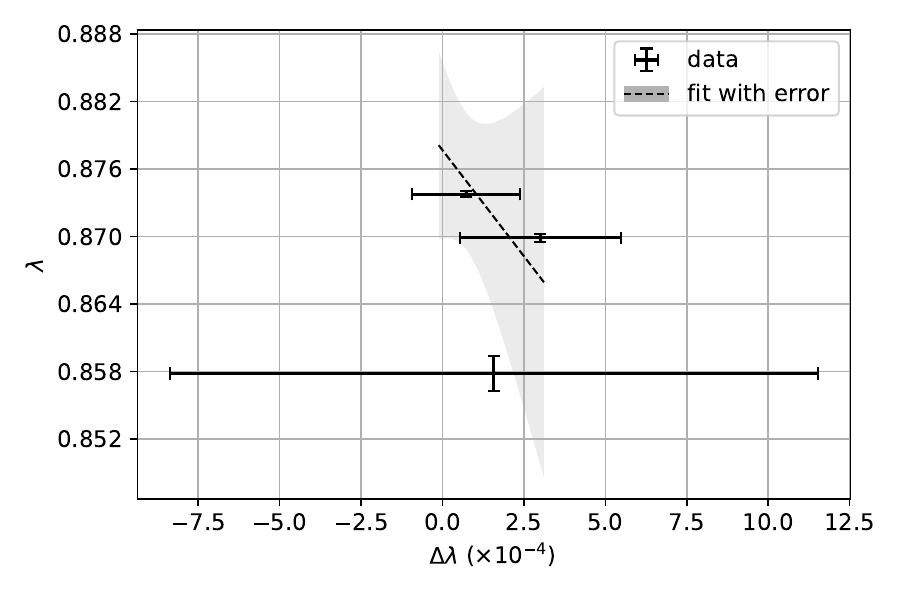}
    \mysubcaption{$a=0.01$}{fig:LAM7OGEVP0p01}
  \end{minipage}
\hfill
  \begin{minipage}{0.49\textwidth}
    \centering
    \includegraphics[width=\linewidth]{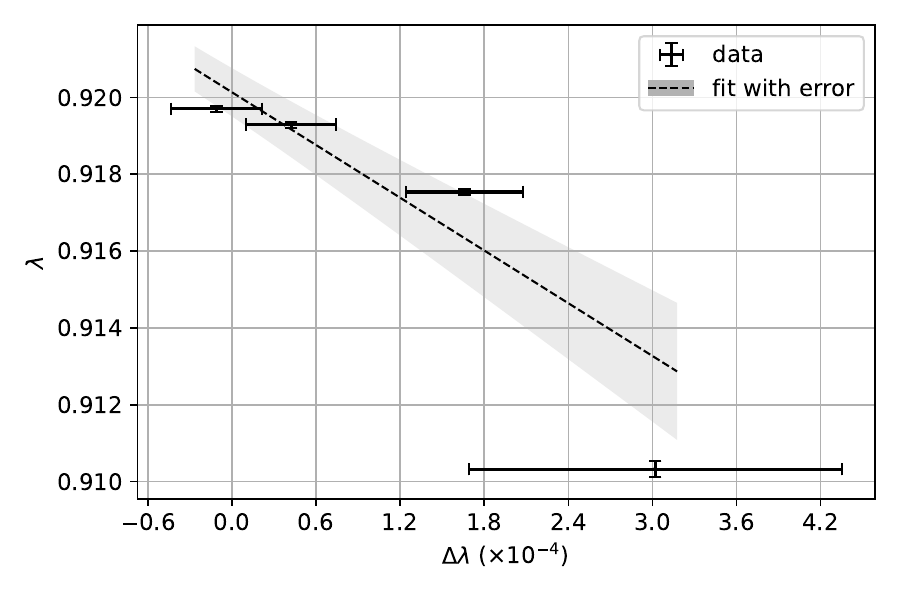}
    \mysubcaption{$a=0.01$}{fig:LAM5OGEVP0p01}
  \end{minipage}
\hfill
  \begin{minipage}{0.49\textwidth}
    \centering
    \includegraphics[width=\linewidth]{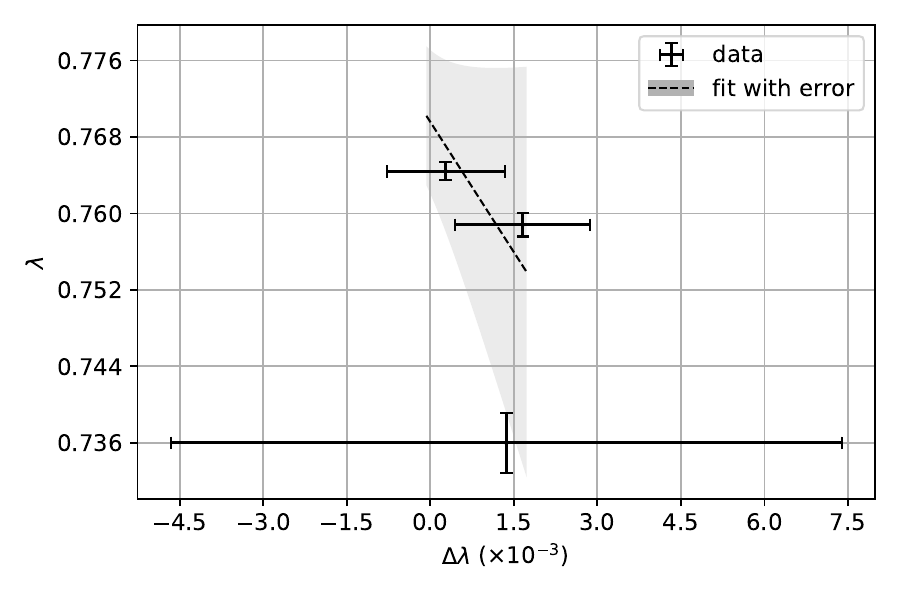}
    \mysubcaption{$a=0.02$}{fig:LAM7OGEVP0p02}
  \end{minipage}
\hfill
  \begin{minipage}{0.49\textwidth}
    \centering
    \includegraphics[width=\linewidth]{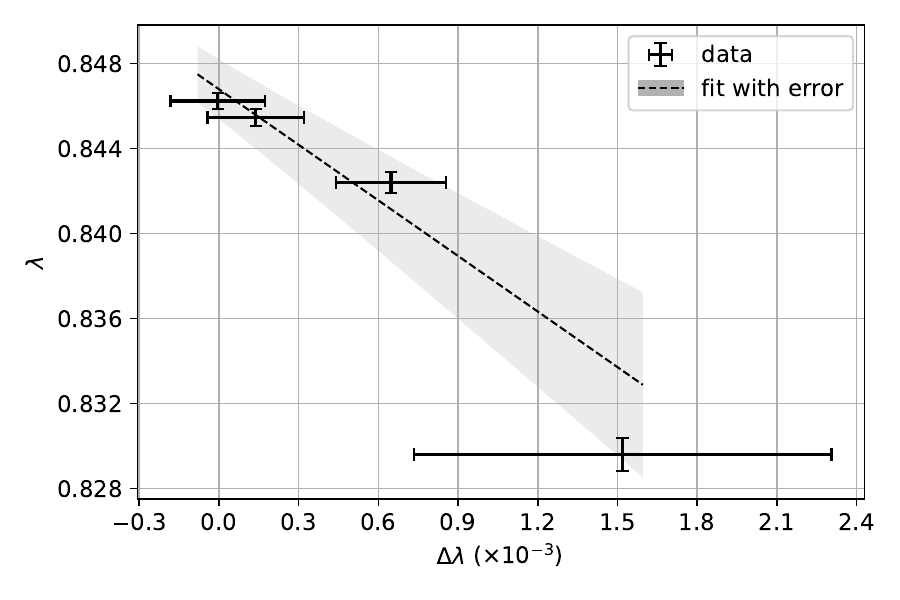}
    \mysubcaption{$a=0.02$}{fig:LAM5OGEVP0p02}
  \end{minipage}
\hfill
  \begin{minipage}{0.49\textwidth}
    \centering
    \includegraphics[width=\linewidth]{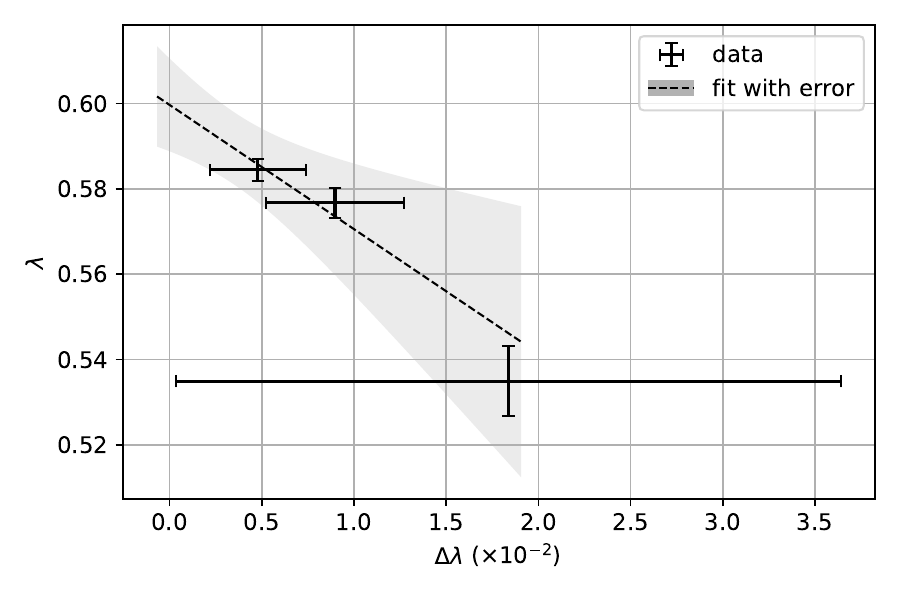}
    \mysubcaption{$a=0.04$}{fig:LAM7OGEVP0p04}
  \end{minipage}
\hfill
  \begin{minipage}{0.49\textwidth}
    \centering
    \includegraphics[width=\linewidth]{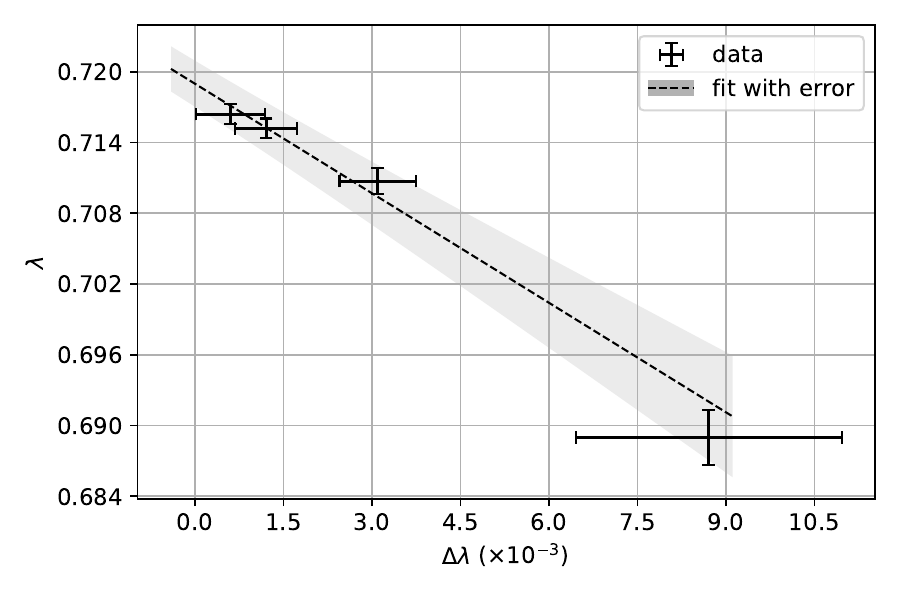}
    \mysubcaption{$a=0.04$}{fig:LAM5OGEVP0p04}
  \end{minipage}
  \caption{ $\lambda$ vs $\Delta\lambda$ plots for the $E_7$ (left column) and $E_5$ (right column)
            excited states at lattice spacings $a=0.01$, $0.02$, and $0.04$.}
  \label{fig:lam7_lam5}
\end{figure}

\begin{figure}[p]
\centering
  \begin{minipage}{0.49\textwidth}
    \centering
    \includegraphics[width=\linewidth]{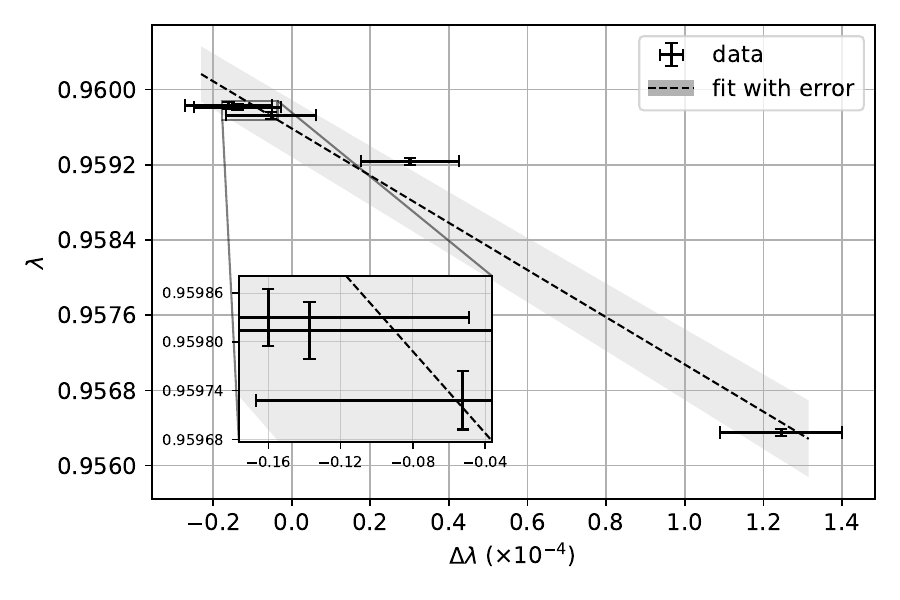}
    \mysubcaption{$a=0.01$}{fig:LAM3OGEVP0p01}
  \end{minipage}
\hfill
  \begin{minipage}{0.49\textwidth}
    \centering
    \includegraphics[width=\linewidth]{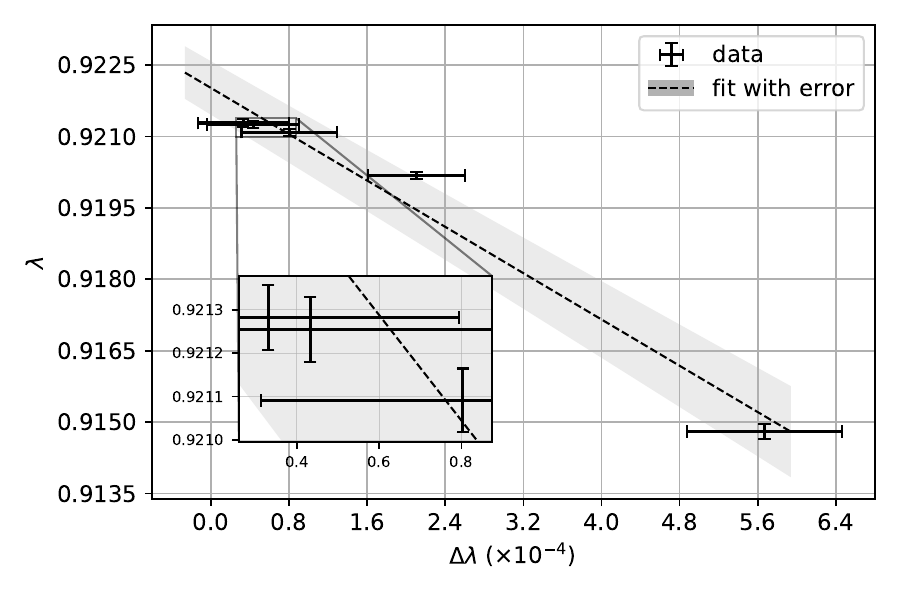}
    \mysubcaption{$a=0.02$}{fig:LAM3OGEVP0p02}
  \end{minipage}
\hfill
  \begin{minipage}{0.49\textwidth}
    \centering
    \includegraphics[width=\linewidth]{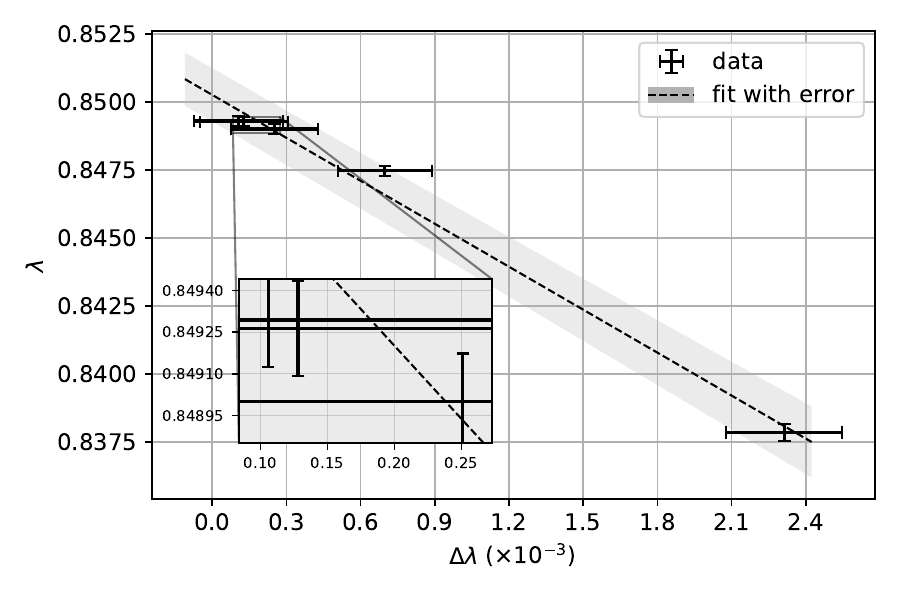}
    \mysubcaption{$a=0.04$}{fig:LAM3OGEVP0p04}
  \end{minipage}
  \caption{Same as Figure~\ref{fig:lam7_lam5}, but for $E_3$.}
  \label{fig:lam3}
\vspace*{2em}
\centering
  \begin{minipage}{0.49\textwidth}
    \centering
    \includegraphics[width=\linewidth]{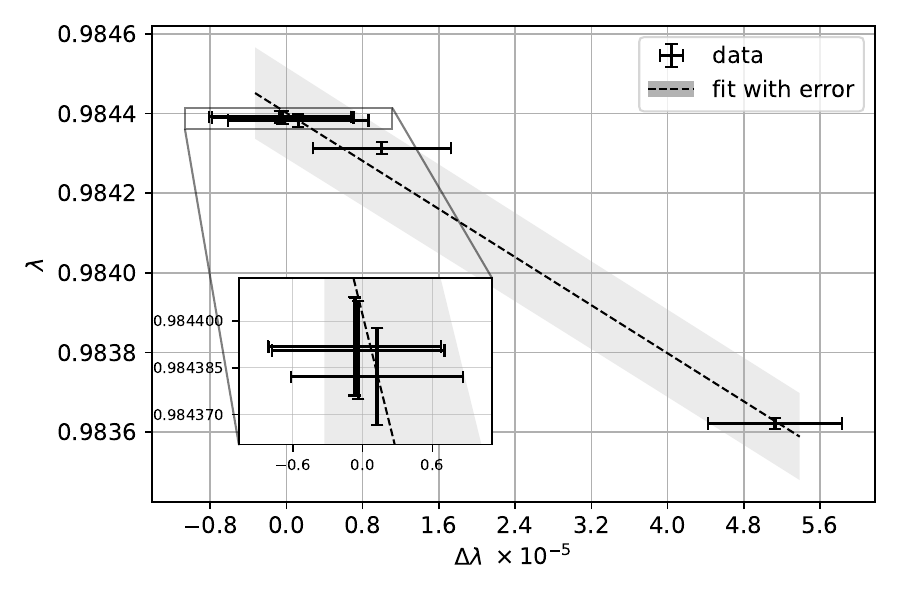}
    \mysubcaption{$a=0.02$}{fig:LAM1withfillOGEVP0p02}
  \end{minipage}
\hfill
  \begin{minipage}{0.49\textwidth}
    \centering
    \includegraphics[width=\linewidth]{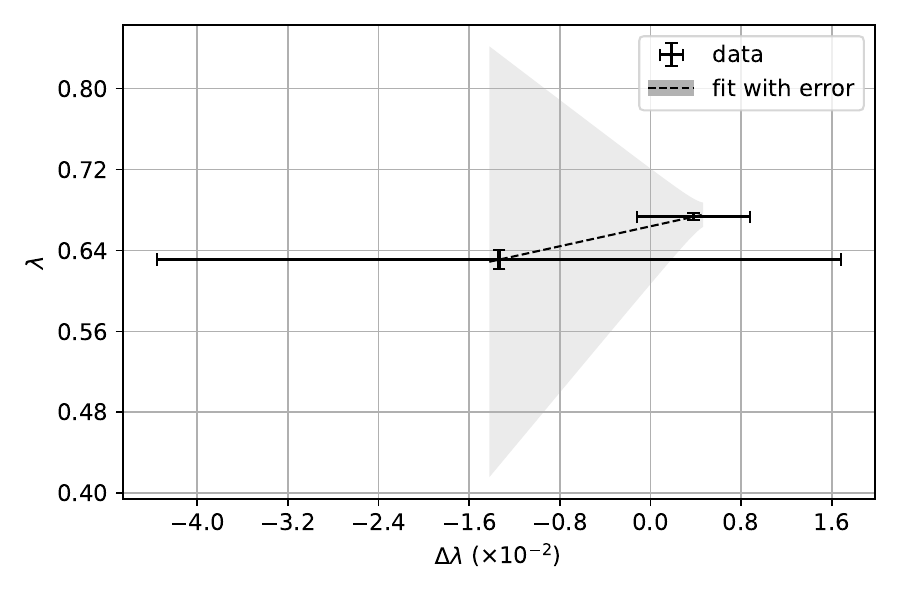}
    \mysubcaption{$a=0.02$}{fig:LAM9withfillOGEVP0p02}
  \end{minipage}
\hfill
  \begin{minipage}{0.49\textwidth}
    \centering
    \includegraphics[width=\linewidth]{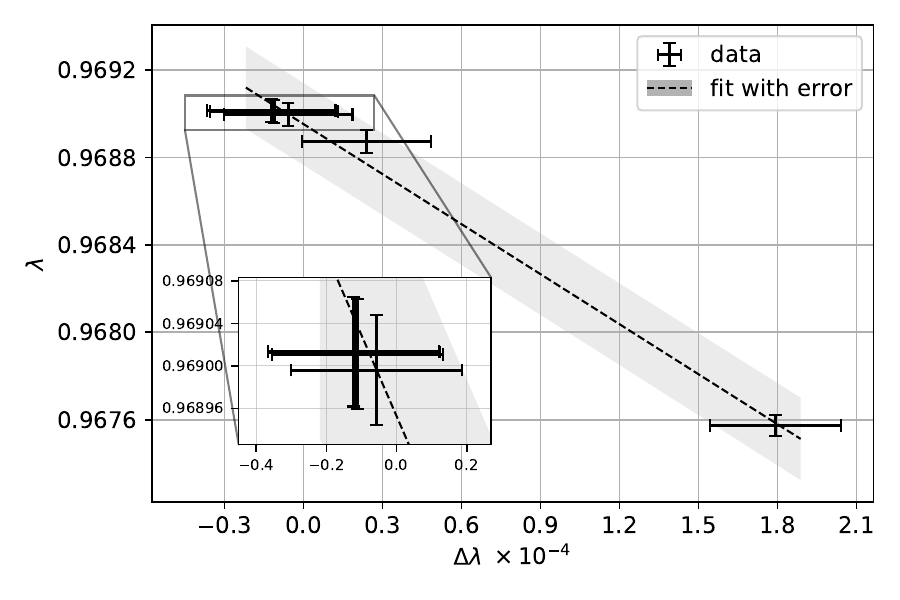}
    \mysubcaption{$a=0.04$}{fig:LAM1withfillOGEVP0p04}
  \end{minipage}
\hfill
  \begin{minipage}{0.49\textwidth}
    \centering
    \includegraphics[width=\linewidth]{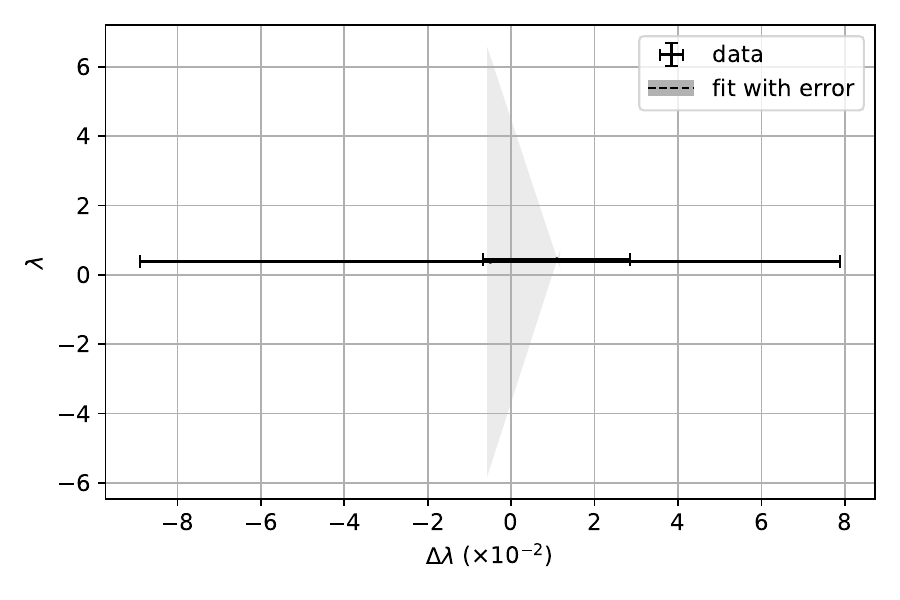}
    \mysubcaption{$a=0.04$}{fig:LAM9withfillOGEVP0p04}
  \end{minipage}
  \caption{Same as Figure~\ref{fig:lam7_lam5}, but for $E_1$ (left) and $E_9$ (right) at lattice spacings $a=0.02$ and $0.04$.
           For the plots at $a=0.01$, see Figs.~\ref{fig:LAM1OGEVP0p01} and \ref{fig:LAM9OGEVP0p01}.}
  \label{fig:lam1_lam9}
\end{figure}

Tables~\ref{tab:E9_E7_lambda}--\ref{tab:E5_E3_E1_lambda} collect $(\lambda,\Delta\lambda)$ data
at each lattice spacing obtained after applying the OGEVP method for anharmonic oscillator.
Figures~\ref{fig:lam7_lam5}--\ref{fig:lam1_lam9} 
are complementary plots to Fig.~\ref{fig:E1andE9} for the eigenvalue-variance extrapolation at each lattice spacing.
We omit the data for $(\lambda,\Delta\lambda)$ for harmonic oscillator.

\begin{table*}[ht]
\centering
\tbl{Comparison of energies extracted using the generalized eigenvalue problem (GEVP) 
     and Operator GEVP (OGEVP) methods
     for harmonic oscillator in single precision (SP) and double precision (DP) 
     for different lattice spacings.\label{tab:precision_comparison_harm}}
{\begin{tabular}{@{}cccccc@{}}
  \toprule
    Lattice spacing $a$ & State &
    GEVP$_{\rm SP}$  &
    GEVP$_{\rm DP}$  &
    OGEVP$_{\rm SP}$ &
    OGEVP$_{\rm DP}$ \\
  \colrule
    \multirow{5}{*}{$0.01$}
    & $E_{1}$ & $1.0029(27)$ & $0.9993(25)$ & $1.0023(25)$ & $1.0003(35)$ \\
    & $E_{3}$ & $3.001(18)$ & $3.027(18)$ & $3.02(80)$ & $3.014(13)$ \\
    & $E_{5}$ & $5.053(53)$ & $5.191(94)$ & $5.04(14)$ & $5.08(16)$ \\
    & $E_{7}$ & $7.98(47)$ & $7.11(95)$ & $7.34(23)$ & $8(4)$ \\
    & $E_{9}$ & $11(1)$ & $9(5)$ & $9(6)$ & $10.65(99)$ \\
  \colrule
    \multirow{5}{*}{$0.02$}
    & $E_{1}$ & $0.9945(26)$ & $0.9994(16)$ & $0.9972(28)$ & $1.0004(22)$ \\
    & $E_{3}$ & $2.984(14)$ & $2.987(12)$ & $2.991(11)$ & $2.98(11)$ \\
    & $E_{5}$ & $4.920(80)$ & $4.995(80)$ & $4.959(59)$ & $5.017(70)$ \\
    & $E_{7}$ & $6.88(33)$ & $7.59(36)$ & $6.96(21)$ & $7.16(24)$ \\
    & $E_{9}$ & $8(2)$ & $11(2)$ & $9.34(38)$ & $9(2)$ \\
  \colrule
    \multirow{5}{*}{$0.04$}
    & $E_{1}$ & $0.99995(198)$ & $0.9979(50)$ & $0.997(10)$ & $0.9985(20)$ \\
    & $E_{3}$ & $3.0015(87)$ & $2.995(10)$ & $2.9997(77)$ & $2.9962(87)$ \\
    & $E_{5}$ & $4.985(74)$ & $4.900(67)$ & $4.981(63)$ & $4.954(73)$ \\
    & $E_{7}$ & $7.02(44)$ & $5.97(76)$ & $7.02(33)$ & $6.65(39)$ \\
    & $E_{9}$ & $12(4)$ & $11(4)$ & $12(35)$ & $8(1)$ \\
  \botrule
\end{tabular}}
\end{table*}

\begin{table*}[ht]
\centering
\tbl{Same as Table~\ref{tab:precision_comparison_harm}, but for anharmonic oscillator. 
     Missing determinations are denoted by ``---''.\label{tab:precision_comparison_aharm}}
{\begin{tabular}{@{}cccccc@{}}
  \toprule
    $a$ & State & GEVP$_{\rm SP}$ & GEVP$_{\rm DP}$ &
    OGEVP$_{\rm SP}$ & OGEVP$_{\rm DP}$ \\
  \colrule
    \multirow{5}{*}{0.01}
    & $E_{1}$ & $0.7871(13)$ & $0.7889(15)$ & $0.7887(35)$ & $0.7976(52)$ \\
    & $E_{3}$ & $4.105(22)$ & $4.1132(91)$ & $4.110(21)$ & $4.125(31)$ \\
    & $E_{5}$ & $8.22(25)$ & $8.32(62)$  & $8.306(74)$ & $8.324(68)$ \\
    & $E_{7}$ & $14(2)$ & $13(8)$ & $13.56(27)$ & $13.05(89)$ \\
    & $E_{9}$ & --- & $24(2)$ & --- & $20(7)$ \\
  \colrule
    \multirow{5}{*}{0.02}
    & $E_{1}$ & $0.7896(21)$ & $0.78748(80)$ & $0.7940(36)$ & $0.7861(58)$ \\
    & $E_{3}$ & $4.104(56)$ & $4.098(11)$ & $4.086(22)$ & $4.060(31)$ \\
    & $E_{5}$ & $8.525(64)$ & $8.43(20)$ & $8.230(69)$ & $8.315(83)$ \\
    & $E_{7}$ & $14.74(56)$ & $14(2)$ & $12(1)$ & $13.10(49)$ \\
    & $E_{9}$ & --- & $26(6)$ & --- & $21(4)$ \\
  \colrule
    \multirow{5}{*}{0.04}
    & $E_{1}$ & $0.7849(23)$ & $0.7873(13)$ & $0.7785(40)$ & $0.7885(48)$ \\
    & $E_{3}$ & $4.0907(77)$ & $4.093(16)$ & $4.056(23)$ & $4.056(29)$ \\
    & $E_{5}$ & $8.698(92)$ & $8.63(23)$ & $8.206(82)$ & $8.248(68)$ \\
    & $E_{7}$ & $16.28(86)$ & $17(2)$ & $13.07(72)$ & $12.78(45)$ \\
    & $E_{9}$ & --- & $49(633)$ & --- & $23(58)$ \\
  \botrule
\end{tabular}}
\end{table*}

\begin{table*}[ht]
\centering
\tbl{$\lambda$, corresponding $\Delta\lambda$, and linear extrapolation results
      for the $E_9$ and $E_7$ states at each $a$ and $N_{\mathrm{OP}}$.
      Numbers in parentheses denote the statistical uncertainty in the last quoted digits.
     Rows with ``Ext.'' are results with the linear eigenvalue-variance extrapolation.
     The values in the $\Delta\lambda$ columns are given in units of $10^{-3}$.\label{tab:E9_E7_lambda}}
{\begin{tabular}{@{}cc%
                S[table-format=-1.4(2)]%
                S[table-format=-2.2(2)]%
                S[table-format=-1.5(2)]%
                S[table-format=-2.2(2)]@{}}
  \toprule
    \multirow{2}{*}{$a$} & \multirow{2}{*}{$N_{\rm op}$}
    & \multicolumn{2}{c}{$\mathbf{E_9}$}
    & \multicolumn{2}{c}{$\mathbf{E_7}$} \\
    \cmidrule{3-6}
    & & {$\lambda$} & {$\Delta\lambda\;(\times10^{-3})$}
      & {$\lambda$} & {$\Delta\lambda\;(\times10^{-3})$} \\
  \colrule
    0.01 & 4 &             &             & 0.8579(15)   & 0.16(99) \\
         & 5 & 0.8023(62)  & 1.2(52)     & 0.8699(4)    & 0.30(25) \\
         & 6 & 0.8188(19)  & -0.3(15)    & 0.8738(3)    & 0.07(17) \\
         & {Ext.} & 0.816(59) & 0 & 0.8777(78) & 0 \\
  \colrule
    0.02 & 4 &             &             & 0.736(3)     & 1.4(60) \\
         & 5 & 0.6309(96) & -13(30)     & 0.7588(12)   & 1.7(12) \\
         & 6 & 0.6732(36)  & 3.8(50)     & 0.76441(95)  & 0.3(11) \\
         & {Ext.} & 0.664(57) & 0 & 0.7696(75) & \\
  \colrule
    0.04 & 4 &             &             & 0.5349(82)   & 18(18) \\
         & 5 & 0.379(21)   & -5(84)      & 0.5768(35)   & 9.0(38) \\
         & 6 & 0.4459(67)  & 11(18)      & 0.5845(26)   & 4.8(26) \\
         & {Ext.} & 0.4(4.1) & 0 & 0.5997(109) & 0 \\
  \botrule
\end{tabular}}
\end{table*}

\begin{table*}[ht]
\centering
\tbl{Same as Table~\ref{tab:E9_E7_lambda}, but for the $E_5$, $E_3$, and $E_1$ states.
    The values in the $\Delta\lambda$ columns are given in units of $10^{-4}$.\label{tab:E5_E3_E1_lambda}}
{\begin{tabular}{@{}cc%
                S[table-format=-1.5(2)]%
                S[table-format=-2.3(3)]%
                S[table-format=-1.5(2)]%
                S[table-format=-2.4(3)]@{}}
  \toprule
    \multirow{2}{*}{$a$} & \multirow{2}{*}{$N_{\rm op}$}
    & \multicolumn{2}{c}{$\mathbf{E_5}$}
    & \multicolumn{2}{c}{$\mathbf{E_3}$} \\
    \cmidrule{3-6}
    & & {$\lambda$} & {$\Delta\lambda\;(\times10^{-4})$}
      & {$\lambda$} & {$\Delta\lambda\;(\times10^{-4})$} \\
  \colrule
    0.01 & 2 &            &             & 0.95635(4)  &  1.25(16) \\
         & 3 & 0.91032(21)&  3.0(13)    & 0.95924(4)  &  0.30(12) \\
         & 4 & 0.91754(8) &  1.66(42)   & 0.95973(4)  & -0.05(11) \\
         & 5 & 0.91929(8) &  0.42(32)   & 0.95981(4)  & -0.14(11) \\
         & 6 & 0.91970(8) & -0.11(33)   & 0.95983(4)  & -0.16(11) \\
         & {Ext.} & 0.92013(63) & 0 & 0.95959(30) & 0 \\
  \colrule
    0.02 & 2 &            &            & 0.9148(2)   & 5.67(80) \\
         & 3 & 0.82961(78)& 15.2(79)   & 0.92018(8)   & 2.11(50) \\
         & 4 & 0.84241(48)& 6.5(21)    & 0.92109(7)   & 0.80(49) \\
         & 5 & 0.84546(40)& 1.4(18)    & 0.92126(8)   & 0.43(47) \\
         & 6 & 0.84624(39)& -0.0(18)   & 0.92128(8)   & 0.33(47) \\
         & {Ext.} & 0.8468(14) & 0 & 0.92202(56) & 0 \\
  \colrule
    0.04 & 2 &           &           & 0.83785(32) & 23.1(24) \\
         & 3 & 0.6890(23)& 87(23)    & 0.84746(19)& 7.0(19) \\
         & 4 & 0.7107(11)& 31.0(65)  & 0.84900(17)& 2.5(18) \\
         & 5 & 0.7152(8) & 12.1(53)  & 0.84926(17)& 1.3(18) \\
         & 6 & 0.7164(9) & 6.0(58)   & 0.84929(17)& 1.1(18) \\
         & {Ext.} & 0.7190(20) & 0 & 0.85025(98) & 0 \\
  \botrule\\
  \multicolumn{6}{c}{%
  \begin{tabular}{@{}cc%
                  S[table-format=-1.6(2)]%
                  S[table-format=-2.4(3)]@{}}
    \toprule
    \multirow{1}{*}{$a$} & \multirow{1}{*}{$N_{\rm op}$}
    & \multicolumn{2}{c}{$\mathbf{E_1}$} \\
    \cmidrule{3-4}
    & & {$\lambda$} & {$\Delta\lambda\;(\times10^{-4})$} \\
   \colrule
    0.01 & 1     & 0.991757(11) &  0.097(16) \\
         & 2     & 0.992108(12) & -0.005(15) \\
         & 3     & 0.992144(12) & -0.029(15) \\
         & 4     & 0.992148(12) & -0.034(15) \\
         & 5     & 0.992149(12) & -0.035(15) \\
         & 6     & 0.992149(12) & -0.035(15) \\
         & {Ext.} & 0.99206(5)  & 0 \\
   \colrule
    0.02 & 1     & 0.983622(14) &  0.513(70) \\
         & 2     & 0.984313(15) &  0.100(72) \\
         & 3     & 0.984382(16) &  0.012(74) \\
         & 4     & 0.984391(16) & -0.004(74) \\
         & 5     & 0.984392(16) & -0.007(74) \\
         & 6     & 0.984392(16) & -0.007(74) \\
         & {Ext.} & 0.98440(11) & 0 \\
   \colrule
    0.04 & 1     & 0.967575(49) &  1.79(25) \\
         & 2     & 0.968873(52) &  2.4(2) \\
         & 3     & 0.968996(52) & -0.06(24) \\
         & 4     & 0.969011(52) & -0.11(24) \\
         & 5     & 0.969013(52) & -0.12(24) \\
         & 6     & 0.969013(52) & -0.12(24) \\
         & {Ext.} & 0.96895(19) & 0 \\
  \botrule
  \end{tabular}}\\
\end{tabular}}
\end{table*}

\clearpage
\bibliographystyle{ws-ijmpa}
\bibliography{Refs_PGEVP}

\end{document}